\documentclass[twocolumn,superscriptaddress,prl]{revtex4-2}
\usepackage{amsmath}
\usepackage{amssymb}
\usepackage{graphicx}
\usepackage{bm}
\usepackage{color}  
\usepackage{hyperref}
\hypersetup{colorlinks=false}

\begin{document}

\title{Charge density waves and Fermi-surface reconstruction in the clean overdoped cuprate superconductor {Tl$_2$Ba$_2$CuO$_{6+\delta}$}}

  \author{C. C. Tam}
  \affiliation{H. H. Wills Physics Laboratory, University of Bristol, Bristol BS8 1TL, United Kingdom.}
 \affiliation{Diamond Light Source, Harwell Campus, Didcot OX11 0DE, United Kingdom.}
    
  \author{M. Zhu}
 \affiliation{H. H. Wills Physics Laboratory, University of Bristol, Bristol BS8 1TL, United Kingdom.}

  \author{J. Ayres}
  \affiliation{H. H. Wills Physics Laboratory, University of Bristol, Bristol BS8 1TL, United Kingdom.}

 \author{K. Kummer}
  \affiliation{ESRF, The European Synchrotron, 71 Avenue des Martyrs, CS40220, 38043 Grenoble Cedex 9, France.}

 \author{F. Yakhou-Harris}
  \affiliation{ESRF, The European Synchrotron, 71 Avenue des Martyrs, CS40220, 38043 Grenoble Cedex 9, France.}

 \author{J. R. Cooper}
  \affiliation{Department of Physics, University of Cambridge,
 Madingley Road, Cambridge, CB3 0HE, United Kingdom.}
    
 \author{A. Carrington}
 \affiliation{H. H. Wills Physics Laboratory, University of Bristol, Bristol BS8 1TL, United Kingdom.}
    
  \author{S. M. Hayden}
  \affiliation{H. H. Wills Physics Laboratory, University of Bristol, Bristol BS8 1TL, United Kingdom.}

\begin{abstract}
Hall effect and quantum oscillation measurements on high temperature cuprate superconductors show that underdoped compositions have a small Fermi surface pocket whereas when heavily overdoped, the pocket increases dramatically in size. The origin of this change in electronic structure has been unclear, but may be related to the high temperature superconductivity.   Here we show that the clean overdoped single-layer cuprate Tl$_2$Ba$_2$CuO$_{6+\delta}$ (Tl2201) displays CDW order with a remarkably long correlation length $\xi \approx 200$\;\AA\ which disappears above a hole concentration $p_{\text{CDW}} \approx 0.265$.  We show that the evolution of the electronic properties of Tl2201  as the doping is lowered may be explained by a Fermi surface reconstruction which accompanies the emergence of the CDW below $p_{\text{CDW}}$. Our results demonstrate importance of CDW correlations in understanding the electronic properties of overdoped cuprates.
\end{abstract}

\maketitle

The normal state electronic properties of the hole doped cuprate superconductors evolve dramatically with doping. This evolution, in part, evidences the emergence and disappearance of competing ordering tendencies and their fluctuations, namely those associated with the pseudogap, charge and spin \cite{Keimer15}.  
Fluctuations associated with these orders are enhanced close to their putative quantum critical points and have been conjectured to play an important role in producing the high superconducting transition temperature in these materials.
   
The pseudogap corresponds to a loss of electronic states near the Fermi energy \cite{Timusk1999_TS,Tallon2001_TaLo} and in most, if not all, cuprates occurs for doping $p<p^{\star}=0.19$ \cite{Tallon2001_TaLo,Tallon2020_TSCL}. In this regime a charge density wave (CDW) order is also found \cite{Ghiringhelli2012_GTMB,Chang2012_CBHC,SilvaNeto2013_SAFC, Comin2013_CFYY,Tabis2014_TLTB, Croft2014_CLSB, Thampy2014_TDCS}. 
It is believed that this charge order causes a reconstruction of the Fermi surface. In YBa$_2$Cu$_3$O$_{6+x}$ (YBCO), the CDW induces a sign reversal in the Hall number $n_H$ at low-temperature for doping $p<0.16$ \cite{LeBoeuf2011_LDVS}. At slightly higher doping, $n_H$ in the high-field, low temperature limit $n_H^{\infty}$ undergoes a rapid increase from $p \rightarrow 1 + p$ over a narrow doping range $0.16 < p < 0.20$, which has been suggested to be caused by the closing of the pseudogap \cite{Badoux2016_BTLG,Storey2016}.   In the overdoped cuprate Tl$_2$Ba$_2$CuO$_{6+\delta}$  (Tl2201), a similar $p\rightarrow 1+p$ transition is observed in $n_H^{\infty}$ which onsets at a higher doping $p \simeq 0.25$, where thermodynamic probes suggest that there is no  pseudogap  \cite{Wade1994_WLMC,Fujiwara1990_FKAS,Kambe1993_KYHU,Putzke2021_PBTA}. Furthermore as there are no reports of a CDW in overdoped Tl2201, a universal microscopic origin of the transition in $n_H^{\infty}$ remains to be found.
Here we use Cu-$L_3$ edge resonant inelastic x-ray scattering (RIXS) to observe a CDW in heavily overdoped Tl2201. The doping onset of the CDW coincides with the decrease in $n_H^{\infty}$ in Tl2201 suggesting these phenomena are linked.

The single layer cuprate, Tl2201, is exceptionally electronically clean and is the only hole doped cuprate in which the large $(1+p)$ Fermi surface has been observed by quantum oscillations (QO) \cite{Vignolle2008_VCCF, Bangura2010, Rourke2010_RBBM}.  The shape of the Fermi surface for the most overdoped compositions ($p\ge 0.27$) has been determined by angle-dependent magnetoresistance \cite{Hussey2003_HACM,AbdelJawad2006_AKBC}, QO and angle resolved photoemission spectroscopy (ARPES) \cite{Plate2005_PMEP} and is in excellent agreement with conventional density functional theory \cite{Rourke2010_RBBM}. Previously, there have been no reports of charge or other ordering in this overdoped material.  However  CDWs  have  been  seen  in  other  overdoped  cuprates.  In La$_{2-x}$Sr$_x$CuO$_4$ (LSCO), the CDW has been seen to persist into the overdoped regime  (up to $p=0.21$) \cite{Lin2020_LMMG}, and a second CDW phase, disconnected from the CDW at lower doping, was observed in overdoped (Bi,Pb)$_{2.12}$Sr$_{1.88}$CuO$_{6+\delta}$ (Bi2201)~\cite{Peng2018_PFDM}. In overdoped Bi$_2$Sr$_2$CaCu$_2$O$_{8+\delta}$ (Bi2212) no CDW is seen directly \cite{He2018_HWSL} although there is phonon softening which may be a precursor of CDW formation.

\begin{figure*}[t]
\includegraphics{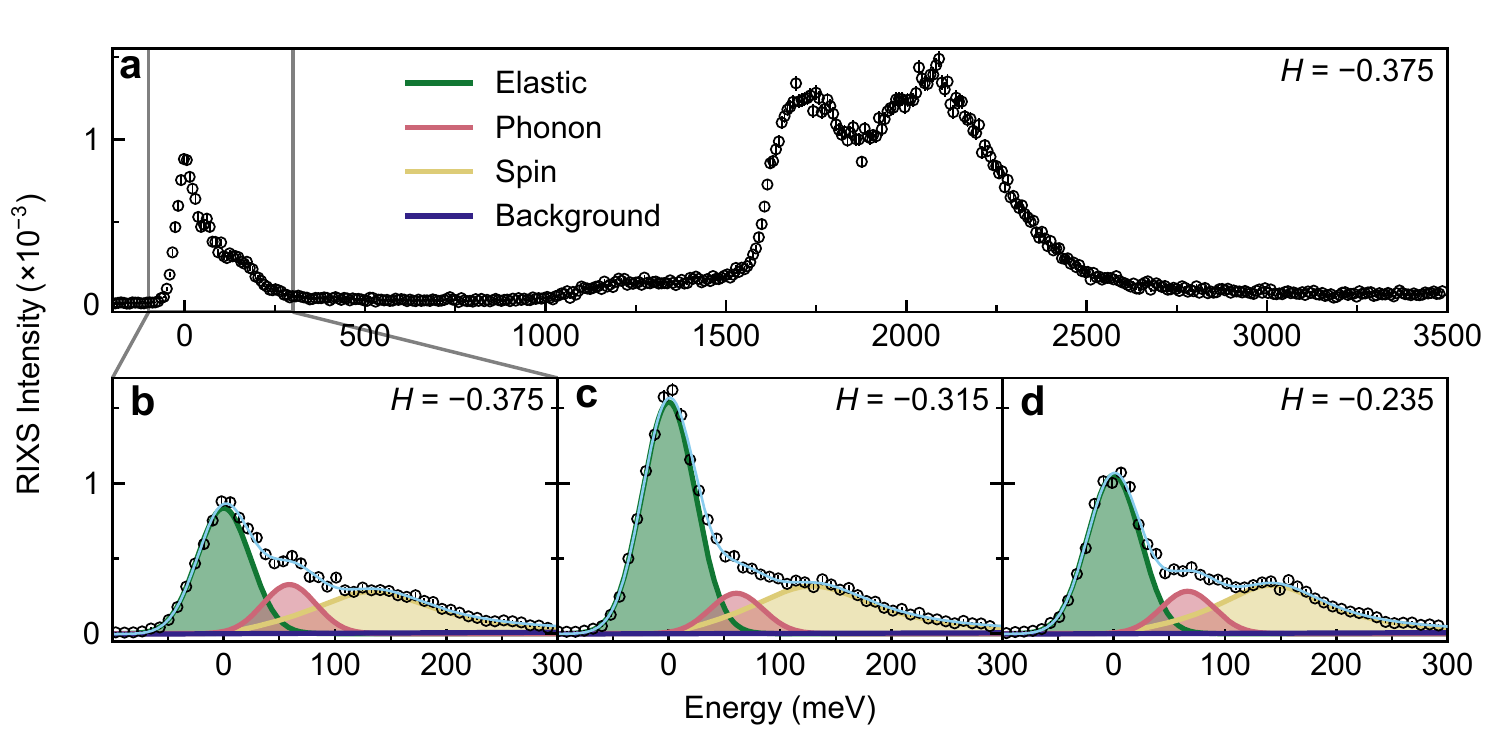}
\caption{\textbf{RIXS spectra.} \textbf{a} RIXS spectra of the $p=0.25$ sample taken at 45\,K showing $dd$ excitations near 2000~meV.   \textbf{b}-\textbf{d} The low-energy region of the RIXS spectra, can be fitted with three components comprising elastic scattering including the CDW, phonon and paramagnon spin excitations (see Methods). The elastic scattering is strongest at $H=-0.315$ where the CDW is present. We label reciprocal space using $\mathbf{Q} = H \mathbf{a}^{\star} + K \mathbf{b}^{\star} + L \mathbf{c}^{\star}$, where $|\mathbf{a}^{\star}|=2 \pi/a$, $a=b=3.85$~\AA, $c=23.1$~\AA, and negative $H$ implies grazing incident x-rays. }
\label{fig:single_spect}
\end{figure*}

\section*{Results}

\noindent\textbf{Charge density wave order in Tl$_2$Ba$_2$CuO$_{6+\delta}$.}  Charge density wave correlations in underdoped superconducting cuprates \cite{Ghiringhelli2012_GTMB,Chang2012_CBHC,SilvaNeto2013_SAFC, Comin2013_CFYY,Tabis2014_TLTB, Croft2014_CLSB, Thampy2014_TDCS} have ordering wavevectors with in-plane components along the CuO bonds i.e. $(\delta,0)$ and $(0,\delta)$, where $0.23 \lesssim \delta \lesssim 0.33$\; r.l.u. In the absence of a large magnetic field or uniaxial stress, a 2D CDW develops below $T_{\textrm{CDW}}$. This state corresponds to a weak anti-correlation of the phase of the CDW in neighbouring CuO$_2$ planes and produces a peak in the scattered intensity at half-integer $L$ positions such as $(\delta,0,2.5)$. In this paper, we label reciprocal space in reciprocal lattice units (r.l.u.) where $\mathbf{Q} = H \mathbf{a}^{\star} + K \mathbf{b}^{\star} + L \mathbf{c}^{\star}$.  We use RIXS (see Methods) to study the charge correlations in three overdoped superconducting samples. 
Our samples have critical temperatures $T_\mathrm{c} = 56$\,K, 45\,K and 22\,K, corresponding to hole doping $p=0.23$, $p=0.25$ and $p=0.28$~\cite{Putzke2021_PBTA}. Fig.~\ref{fig:single_spect}a shows a typical RIXS energy ($E$) dependent spectrum obtained in this experiment showing $dd$ excitations, corresponding to a transition between $2p\rightarrow 3d$ states, near $E=2$\;eV and features at lower energy which may be fitted (see Methods) with an elastic peak, a non-elastic peak near 60\;meV and a broad peak centred at about 150\;meV. We believe the broad 150\;meV peak is due to paramagnon spin fluctuations previously observed in Tl2201 \cite{Tacon2013_TMPM} and the 60\;meV peak to a phonon. 

\begin{figure*}[t]
\includegraphics{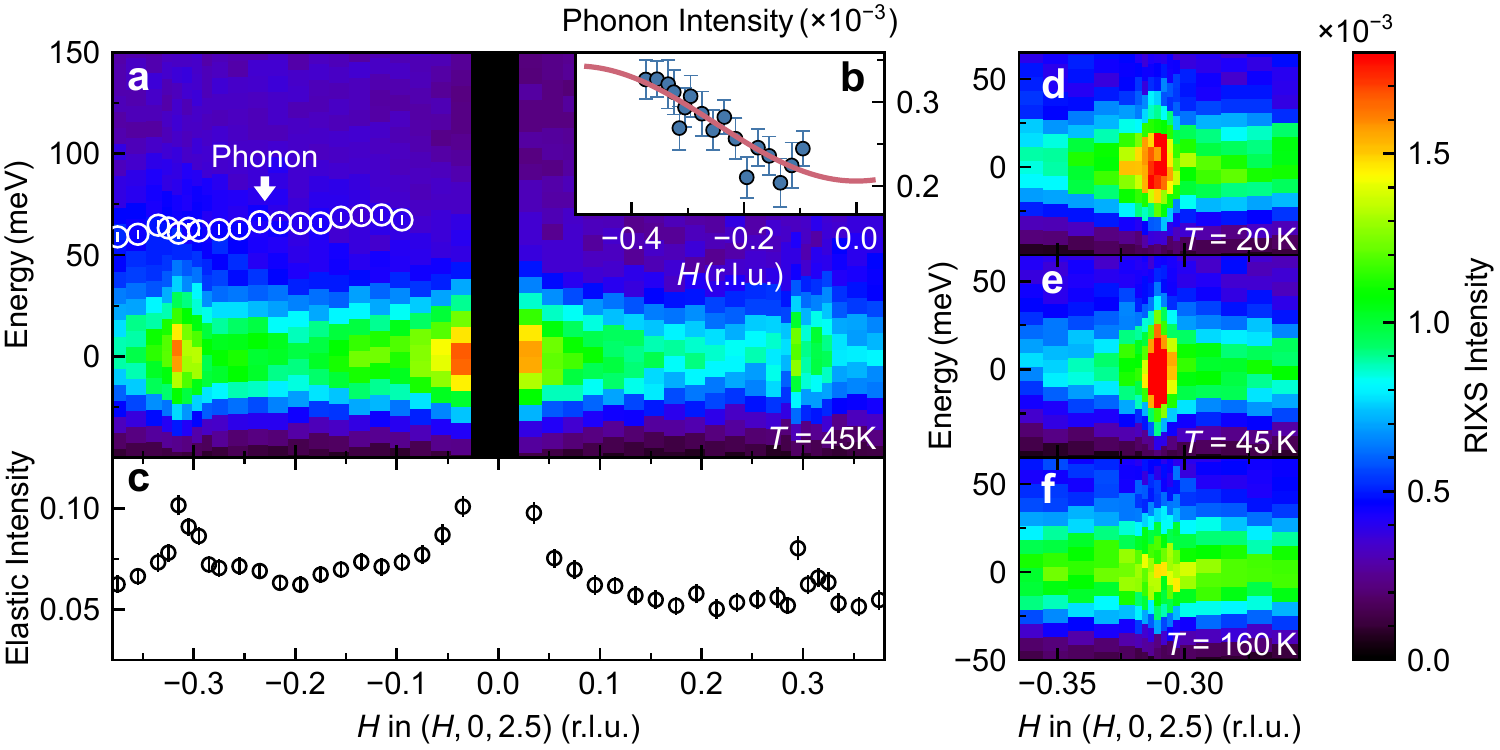}
\caption{\textbf{RIXS intensity maps.} RIXS intensity maps for $p=0.25$ sample taken at 45\,K  \textbf{a} RIXS intensity $H-E$ map over the $H$ range $\pm0.375$. White circles are energies the feature in Fig.\;\ref{fig:single_spect}b-c believed to be a phonon and the intensity of the feature is plotted in the inset in \textbf{b}, along with a line proportional to $\sin^2{(\pi H)}$ in red. \textbf{c} The elastic intensity obtained by integrating the RIXS intensity between $-100 < E < 100$\,meV.  Peaks at $H\approx \pm 0.31$ are due to the charge density wave. \textbf{d}-\textbf{f} are RIXS intensity maps of the peak centred at $H \approx -0.31$ at different temperatures. Intensity maps are normalised to $dd$ excitations.}
\label{fig:multi_spect} 
\end{figure*}

Fig.\;\ref{fig:multi_spect} shows data such as that in Fig.~\ref{fig:single_spect} for the $p=0.25$ ($T_c=45$\;K) sample compiled into $H-E$ colour maps. As mentioned above, the spectra can be fitted to three components. The elastic intensity, obtained by integrating the RIXS intensity between $-100 < E < 100$\,meV (Fig.~\ref{fig:multi_spect}c) is peaked at $H\approx \pm 0.31$. These elastic features are also seen in Fig.\;\ref{fig:multi_spect}a, and we interpret them as charge density wave order as explained below. The elastic scattering is also peaked at $H=0$ due to specular scattering from disorder in the sample surface.  Note that charge scattering is generally enhanced for grazing incident-geometry (denoted by negative $H$) and the vertical polarisation used here.  The energy of the fitted phonon peak is plotted in white circles (Fig.\;\ref{fig:multi_spect}a). Comparing the  dispersion and the $\sin^2{(\pi H)}$ intensity variation (Fig.~\ref{fig:multi_spect}b) to theory~\cite{Devereaux2016_DSWW} we conclude we observe the CuO bond-stretching mode previously seen in Cu-$L$ RIXS on cuprates~\cite{Chaix2017_CGPH}. The $\sin^2{(\pi H)}$ intensity variation can understood in terms of a momentum dependent electron-phonon coupling~\cite{Lin2020_LMMG, Peng_2020}. Our data, which is taken with relatively poor energy resolution, does not exclude a weak softening ($\lesssim$20\%) near $|H| \approx \delta$ \cite{Lin2020_LMMG}. 

\begin{figure*}[t]
\includegraphics{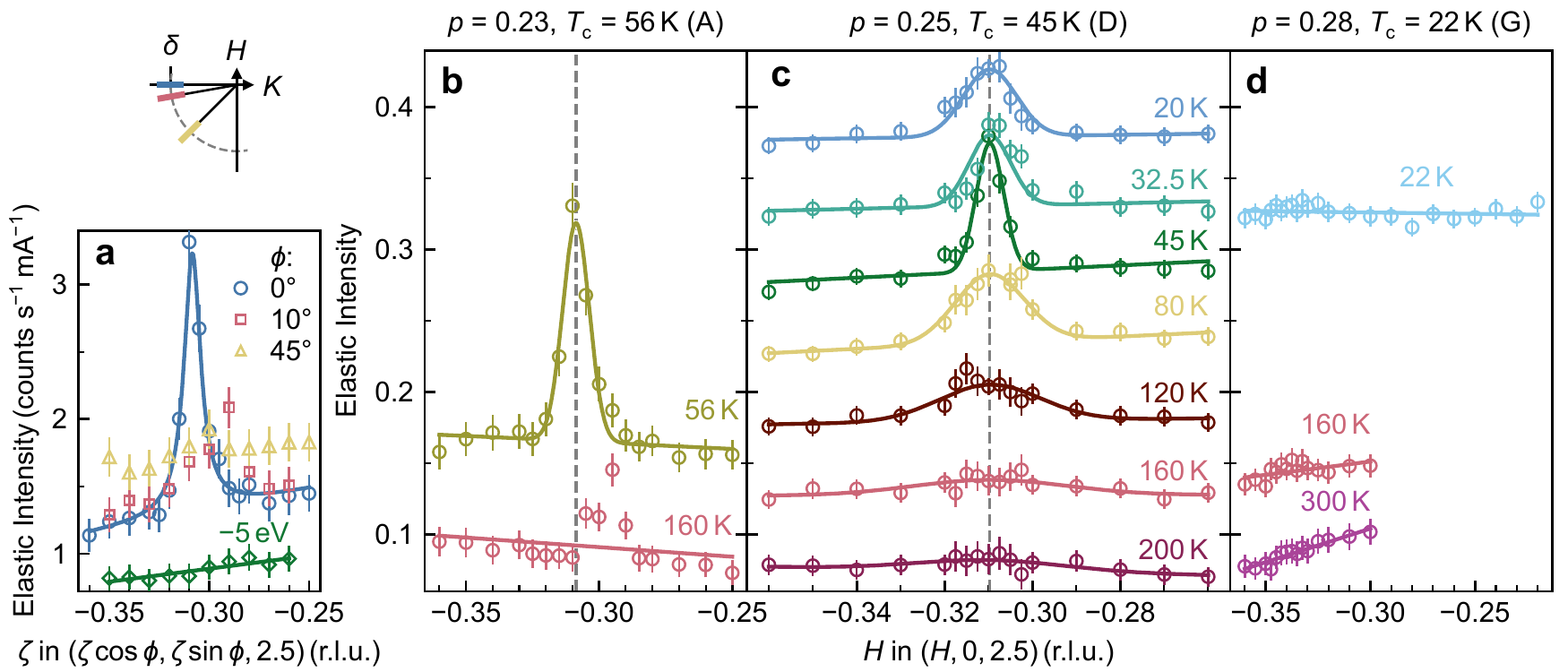}
\caption{\textbf{Wavevector-dependent scans of the CDW order}  Elastic intensities, obtained integrating RIXS intensity over energy in the range $-100< E <100\,$meV, of three samples of Tl2201 at various dopings. \textbf{a} Scans along lines $(\zeta \cos \phi, \zeta \sin \phi, 2.5 )$ showing CDW order occurs near $(-0.31,0,2.5)$ and off-resonance scan with $\phi=0$ for $p=0.23$. Measurements are normalised to the storage ring current. The sketch above indicates the direction of the scans in the $(H,K)$ plane. \textbf{b}-\textbf{d} $(H,0,2.5)$ scans, with intensities normalised to the $dd$ excitations. Measurements at successively lower temperatures in \textbf{c} have been offset by a value of 0.05 dimensionless units per temperature. Data has been fitted to a Gaussian with standard deviation parameter $\sigma$ plus linear background. The grey dashed line marks the centre of the Gaussian peak at $T_\mathrm{c}$, which we have used to determine $\delta$.}
\label{fig:q_cuts}
\end{figure*}

In order to further characterise this CDW feature we investigated other reciprocal space positions and performed non-resonant x-ray scattering. Fig.\;\ref{fig:q_cuts}a shows elastic intensity scans obtained by integrating the RIXS intensity over the energy range $-100<E<100$\,meV for the sample with doping $p=0.23$ at a temperature $T=T_c=56$\;K. The open circles show a scan along $(H,0,2.5)$ yielding a peak at $H=-0.31$\;r.l.u. This peak is not present when scans are made along $(\zeta \cos \phi, \zeta \sin \phi,2.5)$, with $\phi=10^{\circ}$ or 45$^{\circ}$. Nor for non-resonant diffraction when the incident photon energy is lowered by 5\;eV. From this, the data is consistent with a CDW with an in-plane propagation vector of $(\delta,0)$ as observed in other cuprates.   

\begin{figure}[t]
\includegraphics{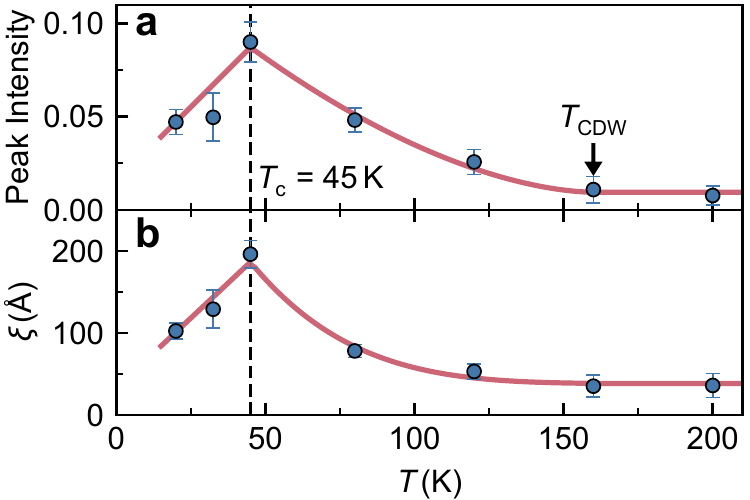}
\caption{\textbf{Temperature dependence of CDW intensity and correlation length.} The CDW peak intensity \textbf{a} and the $a$-axis correlation length \textbf{b} of the $p=0.25$ sample. The dashed line marks $T_\mathrm{c}=45$\,K and the arrow marks $T_\mathrm{CDW}=160\,$K. $\xi(T=T_\mathrm{c}) = 196\pm 17$\,\AA\ and the result of the high temperature $\xi$ fit is $38 \pm 7$\,\AA.  The correlation lengths are $\xi=1/\sigma$ with no correction for instrument resolution and are therefore lower bounds. Error bars are standard deviations determined from least squares fitting.}
\label{fig:T_dependence}
\end{figure}

\begin{figure}[t]
\begin{center}
\includegraphics{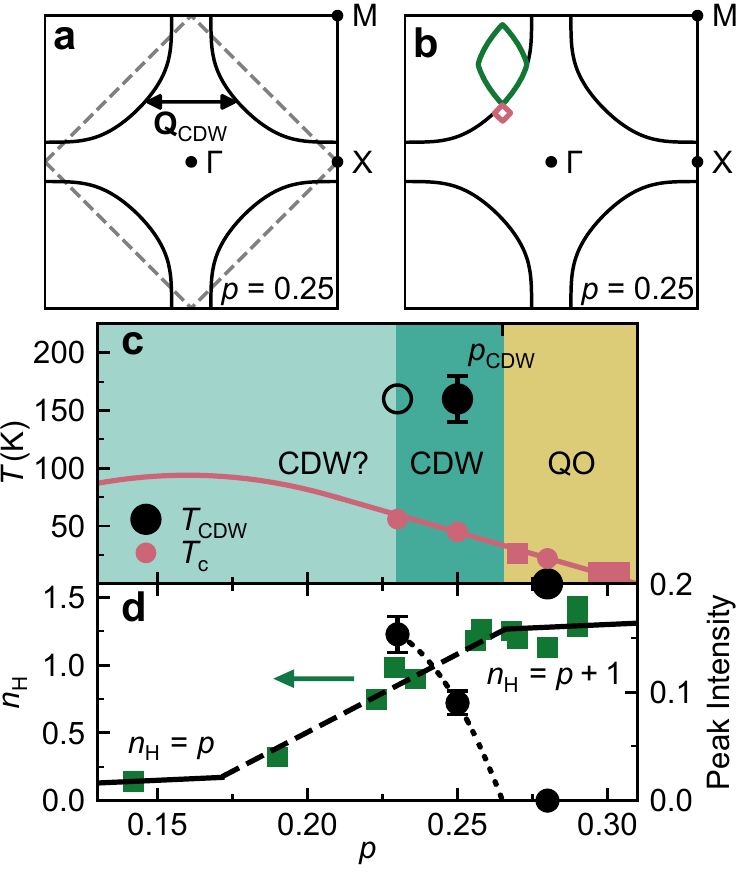}
\end{center}
\caption{\textbf{Charge density wave and Fermi surface reconstruction in Tl$_2$Ba$_2$CuO$_{6+\delta}$}. \textbf{a} 
Fermi surface of Tl2201 (see Methods), with symmetry labels and volume corresponding to 0.25 doped holes (volume is $1+p$). The dashed line is the antiferromagnetic Brillouin zone of the cuprates. \textbf{b}
CDW order may cause a reconstruction yielding electron (red) and hole (green) pockets. 
\textbf{c} Phase diagram of overdoped Tl2201. CDW order is observed for doping $p < p_{\text{CDW}} \approx 0.265$ (dark green region). High-frequency quantum oscillations are observed for $p \ge 0.27$ \cite{Rourke2010_RBBM} (sand coloured region and red squares)  \cite{Rourke2010_RBBM}. CDW existence is undetermined in light green region. CDW onset temperatures $T_{\text{CDW}}$ are shown by black circles, for $p=0.23$ the open circle represents an upper bound. Red line is superconducting $T_\mathrm{c}$ with samples measured here denoted by red filled circles.  \textbf{d} The high-field Hall number~\cite{Putzke2021_PBTA, Mackenzie1996_MJSL} (green squares). Lines of $n_H = p$ and $n_H = p+1$ are marked, with connecting dashed line. CDW peak intensity for $T=T_c$ (black circles). Dotted line is a guide to the eye passing through black circles and $p_{\text{CDW}}$. }
\label{fig:phase_diagram}
\end{figure}

\noindent\textbf{Temperature and doping dependence, correlation length}
Fig.~\ref{fig:q_cuts}b-d, show $H$-scans through the expected position of the CDW for three dopings at various temperatures. At $T \approx T_c$, the samples with dopings $p=0.23$ and $p=0.25$ exhibit peaks centred at $\delta = 0.309 \pm 0.003$ and  $0.310 \pm 0.003$\;r.l.u. respectively, while no peak is observed for the $p=0.28$ sample. We measured a detailed $T$-dependence of the CDW peak in the $p=0.25$ ($T_c=45$\;K) sample (Fig.~\ref{fig:q_cuts}c) to determine the variation the CDW peak intensity $I_{\textrm{CDW}}$ and correlation length $\xi$ (see Fig.\;\ref{fig:T_dependence}), which we define as the reciprocal of the $\sigma$ parameter of the Gaussian fitted to the CDW peak in $H$. The CDW peak intensity and $\xi$ increase below $T=160$\;K, however, there is evidence that they remain finite with $\xi = 38 \pm 7$\,\AA\ to the highest temperature investigated $T=200$\;K. Thus Tl2201 seems to have precursor CDW order seen in other cuprates such as LSCO \cite{Croft2014_CLSB,Miao2021_MFKM} and YBCO \cite{Arpaia2019_ACFV}.  As the temperature is lowered below $T=160$\;K the correlation length increases until $T_c$ is reached, at which point it decreases, suggesting that the superconductivity and the CDW interact.  Similar behaviour is observed in other cuprates \cite{Chang2012_CBHC,Miao2021_MFKM}.  For the $p=0.23$ sample, only two temperatures were measured, $T=T_c=56$\;K and 160\;K. In Fig.~\ref{fig:q_cuts}b, we see that $H=-0.31$ peak has disappeared at 160\,K peak placing an upper bound $T_{\textrm{CDW}}$ of 160\;K for this composition.  We believe the small peak away from $H=-0.31$ is spurious scatting.  For $p=0.28$  (Fig.~\ref{fig:q_cuts}c), no peak is seen even at $T_\mathrm{c}=22$\,K where the signal is expected to be maximal.
From our measurements, we deduce the phase diagram shown in Fig.\;\ref{fig:phase_diagram}c, where the black circles represent $T_{\textrm{CDW}}(p)$, (for $p=0.23$ it is an upper bound) and the critical doping of the CDW is $p_{\textrm{CDW}}=0.265\pm 0.015$, corresponding to $T_c=35 \pm 13$\;K.

\section*{Discussion}

\noindent\textbf{Comparison with the CDW in other cuprates.}
Charge density wave order has been observed in many underdoped ($p<0.16$) cuprates \cite{Ghiringhelli2012_GTMB,Chang2012_CBHC,SilvaNeto2013_SAFC, Comin2013_CFYY,Tabis2014_TLTB, Croft2014_CLSB, Thampy2014_TDCS}, where the ordering wavevectors have been measured as $0.23 < \delta < 0.33$. CDW order has recently been observed by x-ray diffraction in two overdoped materials. In (Bi,Pb)$_{2.12}$Sr$_{1.88}$CuO$_{6+\delta}$ (Bi2201), re-entrant charge order, which is disconnected from the CDW below critical doping is observed \cite{Peng2018_PFDM} and a CDW is observed in overdoped La$_{1.79}$Sr$_{0.21}$CuO$_4$ \cite{Miao2021_MFKM}.  The behaviour of the peak observed in overdoped Bi2201 \cite{Peng2018_PFDM} differs from Tl2201 in that it remains strong up to 250\;K and it is unaffected by the superconductivity.  In contrast, the effect of varying temperature on the peak observed in overdoped LSCO ($p=0.21$) \cite{Miao2021_MFKM} is qualitatively similar to Tl2201 in that its height drops considerably by at temperatures above $T_c$, leaving a small broad component, and the peak height is reduced on entering the superconducting state. The wavevectors of the CDWs in overdoped Tl2201 ($\delta \approx 0.31$) and LSCO ($ \delta \approx 0.24$) are similar to those found in underdoped cuprates ($0.23 < \delta < 0.33$). In the case of LSCO, $ \delta $ does not change much with doping \cite{Miao2021_MFKM} and Tl2201 $\delta \approx 0.31$ is similar to the values found for underdoped YBCO \cite{Ghiringhelli2012_GTMB,Chang2012_CBHC} where $0.31<\delta<0.33$. 

The CDW order in Tl2201 shows a long correlation length for cuprates (in the absence of 3D order induced by magnetic field or strain) of $\xi=196 \pm 17$\;\AA\ for $p=0.25$ at $T=T_c=45\;\textrm{K}$.  In LSCO ($x=0.21$) \cite{Miao2021_MFKM} and YBCO ($p=0.12$)~\cite{Chang2012_CBHC} the maximum correlation lengths are about 80\;\AA\ and 95\;\AA\ respectively. The coherence length is likely limited by quenched disorder induced pinning which may be low in Tl2201, as evidenced by its long electronic mean free path \cite{Rourke2010_RBBM}. 

\noindent\textbf{CDW and the Fermi surface.}
Figs.\;\ref{fig:phase_diagram}a,b shows the Fermi surface of Tl2201 ($p=0.25$) determined from the tight-binding fit of Plat\'e \textit{et al.} \cite{Plate2005_PMEP} based on ARPES, with the chemical potential shifted to give a FS area corresponding to $1+p=1.25$. We place $\mathbf{Q}_{\textrm{CDW}}$ on the same figure in a position where it connects FS states. Johannes and Mazin \cite{Johannes2008_JM} have emphasised that Fermi surface nesting rarely determines CDW order other than in quasi-1D systems. Hence it is not surprising that our $\mathbf{Q}_{\textrm{CDW}}$ does not connect the nested states near the $X$ position.  We note, however, that it does approximately connect Fermi surface states along $\Gamma-M$, which have no $k_z$ dispersion \cite{Hussey2003_HACM} and therefore are nested along $k_z$. However, this condition does not appear to predict $\mathbf{Q}_{\textrm{CDW}}$ in other cuprates.  Our result is also inconsistent with models (e.g.~\cite{Comin2013_CFYY,Pepin2014_PCKM}) that connect charge order with hotspots, where the FS intersects the magnetic zone boundary (dashed line in Fig.\;\ref{fig:phase_diagram}a) determine $\mathbf{Q}_{\textrm{CDW}}$. Thus it appears that determining $\mathbf{Q}_{\textrm{CDW}}$ requires a strong coupling theory of the electron interactions \cite{Kivelson2003_KBFO}.

\begin{figure}
\begin{center}
\includegraphics{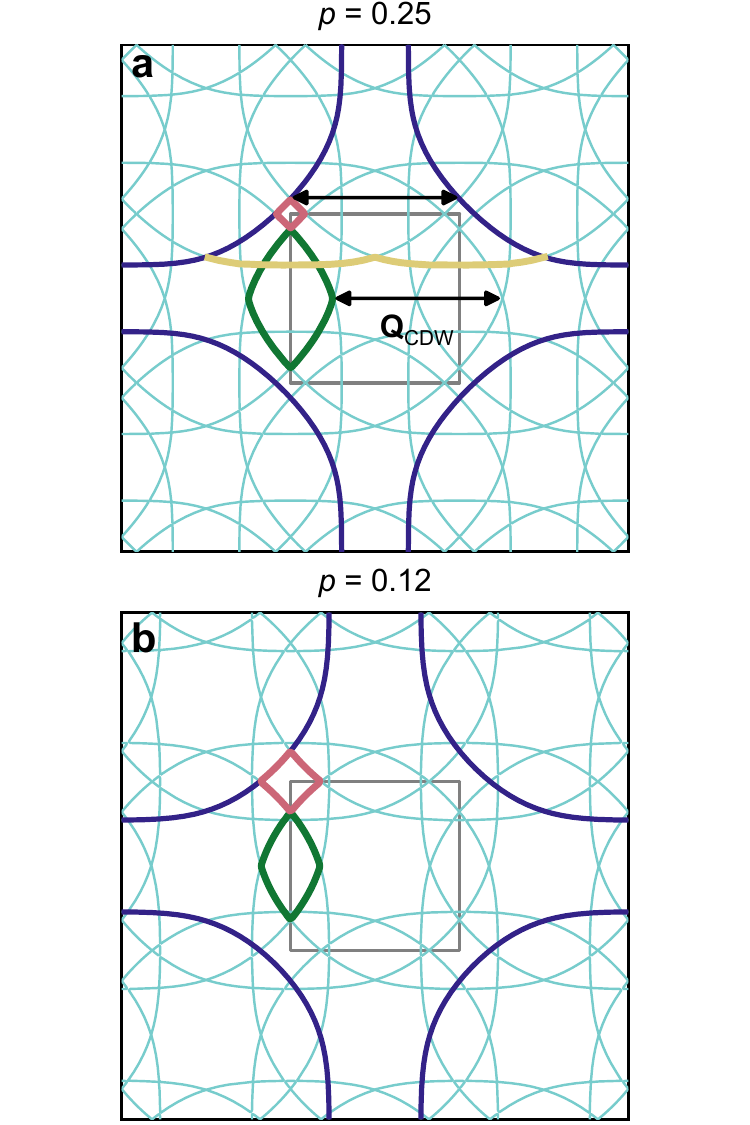} 
\end{center}
\caption{\textbf{Model of Fermi surface reconstruction in Tl2201}. A commensurate $Q=2\pi/3$ has been used. \textbf{a} Shows the reconstruction for $p=0.25$ and \textbf{b} for $p=0.12$. In original unreconstructed FS is show in blue. The reconstructed electron pockets are coloured red, and the hole pockets green.  The open FS sections are coloured sand for $p=0.25$. The reduced Brillouin zone  (BZ) appropriate to the commensurate order is shown by the small grey rectangle in the centre.  The thin turquoise lines show all the contours of the reconstructed FS. For the $p=0.25$ simulation, the hole pockets (2 per BZ) contain 0.054 holes and the electron pockets (1 per BZ) 0.003 electrons.}
\label{fig:FSrecon}
\end{figure}

\noindent\textbf{Hall Effect and Fermi surface reconstruction.}
The appearance of the CDW below $p=0.28$ and its increase in amplitude as the doping is lowered (see Fig.\;\ref{fig:phase_diagram}d) correlates with marked changes in the electronic properties as a function of $p$, in particular $n_H^{\infty}$, as shown in Fig.\;\ref{fig:phase_diagram}d.  Within Boltzmann transport theory, $n_H^{\infty}$ is simply related to the effective number of carriers and so a reduction in $n_H^{\infty}$ could be explained by a reconstruction of the Fermi surface, such as that produced by a CDW.

If the CDW were sufficiently coherent we would expect a reconstruction of the Fermi surface. We have modelled this by adding potentials due to the CDW to a tight binding model of Tl2201 (see Methods).  
The result is shown in Fig.\ \ref{fig:FSrecon}a. Here, for simplicity, we have assumed a commensurate bi-axial order with $|\mathbf{Q}_{\textrm{CDW}}|=1/3$ (compared to $|\mathbf{Q}_{\textrm{CDW}}| \approx 0.31$ in our experiment) and a very small gap (2\,meV).  
To first approximation, the calculation translates the FS by multiples of $\mathbf{Q}_{\textrm{CDW}}$ along the $a$ and $b$ axes,  with gaps appearing where sections of the original FS are connected by $\mathbf{Q}_{\textrm{CDW}}$. In this commensurate case, no gaps appear at the other places where the Fermi surfaces cross and so electrons will follow the trajectory of the original FS. The reconstructed FS therefore consists of electron and hole pockets along with open sections of FS as outlined by the coloured lines in Fig.\ \ref{fig:FSrecon}a.   
 
Although these pockets have a much lower volume than the original unreconstructed Fermi surface, in the limit that the CDW gap is small, calculations \cite{Carrington2021} show that $n_H$ will not be strongly affected.  However, as the gap grows the electron pocket will shrink and eventually disappear and $n_H$ will be reduced to values $\sim p$. The behaviour is similar to the model of Storey \cite{Storey2016} for an antiferromagnetic (pseudogap) reconstruction.  Hence, the evolution of $n_H^{\infty}$ with $p$ in Tl2201 (Fig.\ \ref{fig:phase_diagram}d) might be explained by a growth in the CDW gap as $p$ is reduced.  This is supported by the observed growth of the x-ray intensity (Fig.\ \ref{fig:phase_diagram}d) but further x-ray data, at lower $p$, is required to test this proposition.

In YBCO, the occurrence of the CDW with $p$ is accompanied by a change in sign of $n_H$ \cite{LeBoeuf2011_LDVS}, whereas for Tl2201 $n_H$ remains positive for all $p$ \cite{Putzke2021_PBTA}.  In Fig.\ \ref{fig:FSrecon}b we show a similar calculation where the Fermi-level has been adjusted to give $p=0.12$.  It can be seen that the $p=0.25$ case has larger hole pockets and smaller electron pockets compared to the $p=0.12$ case. In underdoped ($p=0.12$) YBCO, the hole pocket volume is further reduced by the emergence of the pseudogap \cite{Allais2014_AlCS}. The change in sign of $n_H$ in YBCO is explained by the electron pocket contribution being larger than the hole pocket (see Methods).   For Tl2201 calculations \cite{Carrington2021} show that the larger size of the hole pockets and the absence of a pseudogap mean that $n_H$ remains positive even when the CDW gap is large.

\noindent\textbf{Quantum Oscillations.}
The large unreconstructed Fermi sheet centred on the $M$ point in Fig.\;\ref{fig:phase_diagram}b results in quantum oscillations with frequency $F \approx 18$\;kT which have been observed in Tl2201 for $p \ge 0.27$ \cite{Vignolle2008_VCCF,Rourke2010_RBBM}.  The observation of QO is dependent on electrons completing coherent orbits around the Fermi surface.  Thus the disappearance of the 18\;kT frequency for $p < 0.27$ (see Fig.\;\ref{fig:phase_diagram}c) is consistent the CDW causing additional scattering of the orbiting electrons at specific $\mathbf{k}$ vectors. Note that no obvious concomitant increase in the resistivity at $p_{\rm CDW}$ is observed \cite{Putzke2021_PBTA}, so the additional scattering must be quite localised in $\mathbf{k}$.  

The non-observation of lower-frequency QO for $p<0.27$ \cite{Bangura2010,Rourke2010_RBBM} in Tl2201 suggests that the CDW coherence is not sufficient for electrons to complete cyclotron orbits around the smaller reconstructed pockets with the available magnetic field. This is a much stronger criterion than that required to see reconstruction in the Hall effect \cite{Gannot2019_GRK}. 

\noindent\textbf{Conclusion and outlook.}
Spin and charge order or low-frequency fluctuations persist into the overdoped region of the cuprate phase diagram. Low-energy $(\pi,\pi)$ spin fluctuations \cite{Lipscombe2007_LHVM} and charge density wave order \cite{Miao2021_MFKM} are seen in overdoped La$_{2-x}$Sr$_x$CuO$_4$ ($p \approx 0.21$). It is likely that $(\pi,\pi)$ spin fluctuations are also present in overdoped Tl$_2$Ba$_2$CuO$_{6+\delta}$ although they are difficult to measure. Here we show that charge-density wave order with a long correlation length is present in Tl$_2$Ba$_2$CuO$_{6+\delta}$ at higher dopings disappearing at doping $p_{\textrm{CDW}}=0.265 \pm 0.015$, corresponding to $T_c=35 \pm 13$\;K.  The CDW appears in a part of the phase diagram where there is no pseudogap, hence showing the two phenomena are not linked, at least in Tl2201. As in other cuprates \cite{Chang2012_CBHC}, superconductivity in Tl2201 causes the CDW coherence length to shorten and hence the CDW is affected by the superconductivity. 

Overdoped Tl2201 has a number of interesting quasiparticle properties including an increase in strength of a linear-in-$T$ component to the resistivity as doping is reduced from the edge of the superconducting dome \cite{Hussey2008, Putzke2021_PBTA}, quasiparticle scattering rates which are highly anisotropic \cite{AbdelJawad2006_AKBC}, a strong variation of Hall number with doping \cite{Putzke2021_PBTA} and a magnetoresistance which is much larger than predicted by Boltzmann theory and is insensitive to impurities and magnetic field direction \cite{Ayres2021}. We have argued that the CDW order is a potential explanation for the variation of the Hall number with doping. Future work is needed to establish whether the other properties mentioned above can be understood in terms of the spin and charge correlations, and the associated scattering of quasiparticles near the Fermi surface.

\section*{Materials and Methods \label{sec:methods}}

\textbf{Sample growth and preparation.} Single crystals of Tl2201 were grown by a self-flux method, similar to that described in reference~\cite{Tyler1998_Tyle}. The three samples measured roughly $100 \times 100\mu$m, and were hole doped to values of $p$ = 0.23 ($T_c=56$\;K), 0.25 ($T_c=47$\;K) and 0.28 ($T_c=22$\;K), or samples A, D, and G respectively. Sample G was annealed in pure oxygen at 325$^\circ$\;C. Samples A and D were annealed in 0.1 \% oxygen in argon at 550$^\circ$C and 500$^\circ$C respectively. Samples were annealled for up to 64 hours and quenched by quickly removing them from the furnace and placing on a copper block. $T_\mathrm{c}$ was determined from the mid-point of a.c. susceptibility superconducting transition. Doping values were determined from $T_\mathrm{c}$ using the relation given in Ref.\ \cite{Putzke2021_PBTA}. Tl2201 is tetragonal, with approximate lattice parameters $a=b=3.85\,$\AA\ and $c=23.1\,$\AA. The samples had mirror-like surfaces as grown and were not cleaved.

\textbf{Resonant inelastic x-ray scattering.}
RIXS was carried out at the ID32 beamline at the ESRF, Grenoble, France. Using x-ray absorption spectroscopy (XAS), the incident x-ray energy was tuned to the Cu-$L_3$ edge at 931.6\,eV. Data were collected with photons polarised linearly, vertical to the scattering plane (LV), in the low resolution/high throughput configuration of the instrument. The energy resolution was further relaxed from 42 meV to 48 meV by opening the exit slit in order to observe weak elastic scattering. The UB matrix (sample orientation) was determined on ID32 using diffraction from the (002) and (103) Bragg reflections. $L$ was fixed to a half-integer value to maximise the CDW response. Scans in the $(H,0,L)$ plane were made changing sample rotation $\theta$ and detector position $2\theta$ in the horizontal plane. Except for those shown in Fig.~\ref{fig:q_cuts}a, data was normalised by the orbital excitations, obtained by integrating RIXS spectra in the energy range $1000<E < 3000 \,$meV, resulting in a dimensionless ratio of intensities. Error bars on RIXS intensity and the elastic intensity were obtained by taking advantage of single photon counting at ID32, assuming Poisson statistics.

The low energy excitations of the RIXS spectra, where $0<E<300$\,meV were fitted to three features. The elastic peaks were fitted to a Gaussian with a FWHM fixed to the the instrument resolution. This was used to determine zero energy. An inelastic Gaussian peak with FWHM fixed to instrument resolution was fitted to the CuO stretching phonon where present (discussed more in main text).  The magnetic paramagnon excitation was fitted to a damped harmonic oscillator that was numerically convoluted with a Gaussian with FWHM set to instrument resolution. To account any higher energy excitations, a linear background was used. 

\textbf{Fermi surface reconstruction}
The simulations of Fermi-surface reconstruction in Fig.\ \ref{fig:FSrecon} was based on the ARPES derived tight-binding parameterisation of the FS of Tl2201 \cite{Plate2005_PMEP}; $\varepsilon(K)=\mu+\frac{t_1}{2}(\cos k_x+\cos k_y)+t_2 \cos k_x \cos k_y+\frac{t_3}{2}(\cos 2k_x+ \cos 2k_y)+ \frac{t_4}{2}(\cos 2k_x\cos k_y + \cos k_x \cos 2k_y)+t_5 \cos 2k_x\cos 2k_y$, with $t_1=-0.725$, $t_2=0.302$, $t_3=0.0159$, $t_4=-0.0805$ and $t_5=0.0034$; for $p=0.25$ $\mu=0.2382$ and $p=0.12$ $\mu=0.1998$. All energies are expressed in eV with lattice parameter $a=1$. Small CDW potentials with $V_0=0.002$, $\mathbf{Q}_{\text{CDW}}= (2\pi/3,0)$ and $\mathbf{Q}_{\text{CDW}}= (0,2\pi/3)$ were introduced using the method described in Ref.\ \cite{Harrison2011}, but with a $\mathbf{k}$-independent potential.  This translates copies of the FS by multiples of all wavevector $\mathbf{Q}_{\text{CDW}}$ in the $\mathbf{a}$ and $\mathbf{b}$ directions and introduces small gaps on those parts of the FS connected by $\mathbf{Q}_{\text{CDW}}$. In the incommensurate case, which corresponds to our experimental $|\mathbf{Q}_{\text{CDW}}| \approx 0.31$\;r.l.u., small gaps will also be present where the other sections of FS cross.  However, these gaps will generally be small so the electrons will be able to tunnel across in moderate fields and hence the transport will be largely unaffected \cite{Harrison2011}.

In the low field limit, $n_H=(B/e) \sigma_{xx}^2/\sigma_{xy}$, where  $\sigma_{xx}$ and   $\sigma_{xy}$ are the longitudinal and Hall conductivities respectively and $B$ is the magnetic field.  The sign of $n_H$ is therefore the same as $\sigma_{xy}$, which is the sum of positive and negative contributions from the hole and electron pockets respectively which in turn depend predominantly on the electron mean-free-path on each pocket.  The open sections of the Fermi surface will have only a small contribution to $\sigma_{xy}$ as their contribution is zero if the mean-free-path is isotropic \cite{Ong1991_Ong}. These open sections will however, make a substantial contribution to $\sigma_{xx}$ so will affect the magnitude of $n_H$.

\section{Acknowledgements}
We acknowledge useful discussions with Nigel Hussey, Steve Kivelson and Cyril Proust. Liam Malone contributed to the sample growth. The work was supported by the UK Engineering and Physical Sciences Research Council under grant R130114-101.


\begin{thebibliography}{46}%
\makeatletter
\providecommand \@ifxundefined [1]{%
 \@ifx{#1\undefined}
}%
\providecommand \@ifnum [1]{%
 \ifnum #1\expandafter \@firstoftwo
 \else \expandafter \@secondoftwo
 \fi
}%
\providecommand \@ifx [1]{%
 \ifx #1\expandafter \@firstoftwo
 \else \expandafter \@secondoftwo
 \fi
}%
\providecommand \natexlab [1]{#1}%
\providecommand \enquote  [1]{``#1''}%
\providecommand \bibnamefont  [1]{#1}%
\providecommand \bibfnamefont [1]{#1}%
\providecommand \citenamefont [1]{#1}%
\providecommand \href@noop [0]{\@secondoftwo}%
\providecommand \href [0]{\begingroup \@sanitize@url \@href}%
\providecommand \@href[1]{\@@startlink{#1}\@@href}%
\providecommand \@@href[1]{\endgroup#1\@@endlink}%
\providecommand \@sanitize@url [0]{\catcode `\\12\catcode `\$12\catcode
  `\&12\catcode `\#12\catcode `\^12\catcode `\_12\catcode `\%12\relax}%
\providecommand \@@startlink[1]{}%
\providecommand \@@endlink[0]{}%
\providecommand \url  [0]{\begingroup\@sanitize@url \@url }%
\providecommand \@url [1]{\endgroup\@href {#1}{\urlprefix }}%
\providecommand \urlprefix  [0]{URL }%
\providecommand \Eprint [0]{\href }%
\providecommand \doibase [0]{https://doi.org/}%
\providecommand \selectlanguage [0]{\@gobble}%
\providecommand \bibinfo  [0]{\@secondoftwo}%
\providecommand \bibfield  [0]{\@secondoftwo}%
\providecommand \translation [1]{[#1]}%
\providecommand \BibitemOpen [0]{}%
\providecommand \bibitemStop [0]{}%
\providecommand \bibitemNoStop [0]{.\EOS\space}%
\providecommand \EOS [0]{\spacefactor3000\relax}%
\providecommand \BibitemShut  [1]{\csname bibitem#1\endcsname}%
\let\auto@bib@innerbib\@empty
\bibitem [{\citenamefont {Keimer}\ \emph {et~al.}(2015)\citenamefont {Keimer},
  \citenamefont {Kivelson}, \citenamefont {Norman}, \citenamefont {Uchida},\
  and\ \citenamefont {Zaanen}}]{Keimer15}%
  \BibitemOpen
  \bibfield  {author} {\bibinfo {author} {\bibfnamefont {B.}~\bibnamefont
  {Keimer}}, \bibinfo {author} {\bibfnamefont {S.~A.}\ \bibnamefont
  {Kivelson}}, \bibinfo {author} {\bibfnamefont {M.~R.}\ \bibnamefont
  {Norman}}, \bibinfo {author} {\bibfnamefont {S.}~\bibnamefont {Uchida}},\
  and\ \bibinfo {author} {\bibfnamefont {J.}~\bibnamefont {Zaanen}},\
  }\bibfield  {title} {\bibinfo {title} {From quantum matter to
  high-temperature superconductivity in copper oxides},\ }\href
  {https://doi.org/10.1038/nature14165} {\bibfield  {journal} {\bibinfo
  {journal} {Nature}\ }\textbf {\bibinfo {volume} {518}},\ \bibinfo {pages}
  {179} (\bibinfo {year} {2015})}\BibitemShut {NoStop}%
\bibitem [{\citenamefont {Timusk}\ and\ \citenamefont
  {Statt}(1999)}]{Timusk1999_TS}%
  \BibitemOpen
  \bibfield  {author} {\bibinfo {author} {\bibfnamefont {T.}~\bibnamefont
  {Timusk}}\ and\ \bibinfo {author} {\bibfnamefont {B.}~\bibnamefont {Statt}},\
  }\bibfield  {title} {\bibinfo {title} {The pseudogap in high-temperature
  superconductors: an experimental survey},\ }\href
  {https://doi.org/10.1088/0034-4885/62/1/002} {\bibfield  {journal} {\bibinfo
  {journal} {Rep. Prog. Phys}\ }\textbf {\bibinfo {volume} {62}},\ \bibinfo
  {pages} {61} (\bibinfo {year} {1999})}\BibitemShut {NoStop}%
\bibitem [{\citenamefont {Tallon}\ and\ \citenamefont
  {Loram}(2001)}]{Tallon2001_TaLo}%
  \BibitemOpen
  \bibfield  {author} {\bibinfo {author} {\bibfnamefont {J.}~\bibnamefont
  {Tallon}}\ and\ \bibinfo {author} {\bibfnamefont {J.}~\bibnamefont {Loram}},\
  }\bibfield  {title} {\bibinfo {title} {The doping dependence of {$T^{\star}$}
  – {What} is the real high-${T}_c$ phase diagram?},\ }\href
  {https://doi.org/https://doi.org/10.1016/S0921-4534(00)01524-0} {\bibfield
  {journal} {\bibinfo  {journal} {Physica C: Superconductivity}\ }\textbf
  {\bibinfo {volume} {349}},\ \bibinfo {pages} {53} (\bibinfo {year}
  {2001})}\BibitemShut {NoStop}%
\bibitem [{\citenamefont {Tallon}\ \emph {et~al.}(2020)\citenamefont {Tallon},
  \citenamefont {Storey}, \citenamefont {Cooper},\ and\ \citenamefont
  {Loram}}]{Tallon2020_TSCL}%
  \BibitemOpen
  \bibfield  {author} {\bibinfo {author} {\bibfnamefont {J.~L.}\ \bibnamefont
  {Tallon}}, \bibinfo {author} {\bibfnamefont {J.~G.}\ \bibnamefont {Storey}},
  \bibinfo {author} {\bibfnamefont {J.~R.}\ \bibnamefont {Cooper}},\ and\
  \bibinfo {author} {\bibfnamefont {J.~W.}\ \bibnamefont {Loram}},\ }\bibfield
  {title} {\bibinfo {title} {Locating the pseudogap closing point in cuprate
  superconductors: Absence of entrant or reentrant behavior},\ }\href
  {https://doi.org/10.1103/physrevb.101.174512} {\bibfield  {journal} {\bibinfo
   {journal} {Phys. Rev. B}\ }\textbf {\bibinfo {volume} {101}},\ \bibinfo
  {pages} {174512} (\bibinfo {year} {2020})}\BibitemShut {NoStop}%
\bibitem [{\citenamefont {Ghiringhelli}\ \emph {et~al.}(2012)\citenamefont
  {Ghiringhelli}, \citenamefont {Tacon}, \citenamefont {Minola}, \citenamefont
  {Blanco-Canosa}, \citenamefont {Mazzoli}, \citenamefont {Brookes},
  \citenamefont {Luca}, \citenamefont {Frano}, \citenamefont {Hawthorn},
  \citenamefont {He}, \citenamefont {Loew}, \citenamefont {Sala}, \citenamefont
  {Peets}, \citenamefont {Salluzzo}, \citenamefont {Schierle}, \citenamefont
  {Sutarto}, \citenamefont {Sawatzky}, \citenamefont {Weschke}, \citenamefont
  {Keimer},\ and\ \citenamefont {Braicovich}}]{Ghiringhelli2012_GTMB}%
  \BibitemOpen
  \bibfield  {author} {\bibinfo {author} {\bibfnamefont {G.}~\bibnamefont
  {Ghiringhelli}}, \bibinfo {author} {\bibfnamefont {M.~L.}\ \bibnamefont
  {Tacon}}, \bibinfo {author} {\bibfnamefont {M.}~\bibnamefont {Minola}},
  \bibinfo {author} {\bibfnamefont {S.}~\bibnamefont {Blanco-Canosa}}, \bibinfo
  {author} {\bibfnamefont {C.}~\bibnamefont {Mazzoli}}, \bibinfo {author}
  {\bibfnamefont {N.~B.}\ \bibnamefont {Brookes}}, \bibinfo {author}
  {\bibfnamefont {G.~M.~D.}\ \bibnamefont {Luca}}, \bibinfo {author}
  {\bibfnamefont {A.}~\bibnamefont {Frano}}, \bibinfo {author} {\bibfnamefont
  {D.~G.}\ \bibnamefont {Hawthorn}}, \bibinfo {author} {\bibfnamefont
  {F.}~\bibnamefont {He}}, \bibinfo {author} {\bibfnamefont {T.}~\bibnamefont
  {Loew}}, \bibinfo {author} {\bibfnamefont {M.~M.}\ \bibnamefont {Sala}},
  \bibinfo {author} {\bibfnamefont {D.~C.}\ \bibnamefont {Peets}}, \bibinfo
  {author} {\bibfnamefont {M.}~\bibnamefont {Salluzzo}}, \bibinfo {author}
  {\bibfnamefont {E.}~\bibnamefont {Schierle}}, \bibinfo {author}
  {\bibfnamefont {R.}~\bibnamefont {Sutarto}}, \bibinfo {author} {\bibfnamefont
  {G.~A.}\ \bibnamefont {Sawatzky}}, \bibinfo {author} {\bibfnamefont
  {E.}~\bibnamefont {Weschke}}, \bibinfo {author} {\bibfnamefont
  {B.}~\bibnamefont {Keimer}},\ and\ \bibinfo {author} {\bibfnamefont
  {L.}~\bibnamefont {Braicovich}},\ }\bibfield  {title} {\bibinfo {title}
  {Long-range incommensurate charge fluctuations in
  {(Y,Nd)Ba}$_2${Cu}$_3${O}$_{6+x}$},\ }\href
  {https://doi.org/10.1126/science.1223532} {\bibfield  {journal} {\bibinfo
  {journal} {Science}\ }\textbf {\bibinfo {volume} {337}},\ \bibinfo {pages}
  {821} (\bibinfo {year} {2012})}\BibitemShut {NoStop}%
\bibitem [{\citenamefont {Chang}\ \emph {et~al.}(2012)\citenamefont {Chang},
  \citenamefont {Blackburn}, \citenamefont {Holmes}, \citenamefont
  {Christensen}, \citenamefont {Larsen}, \citenamefont {Mesot}, \citenamefont
  {Liang}, \citenamefont {Bonn}, \citenamefont {Hardy}, \citenamefont
  {Watenphul}, \citenamefont {v.~Zimmermann}, \citenamefont {Forgan},\ and\
  \citenamefont {Hayden}}]{Chang2012_CBHC}%
  \BibitemOpen
  \bibfield  {author} {\bibinfo {author} {\bibfnamefont {J.}~\bibnamefont
  {Chang}}, \bibinfo {author} {\bibfnamefont {E.}~\bibnamefont {Blackburn}},
  \bibinfo {author} {\bibfnamefont {A.~T.}\ \bibnamefont {Holmes}}, \bibinfo
  {author} {\bibfnamefont {N.~B.}\ \bibnamefont {Christensen}}, \bibinfo
  {author} {\bibfnamefont {J.}~\bibnamefont {Larsen}}, \bibinfo {author}
  {\bibfnamefont {J.}~\bibnamefont {Mesot}}, \bibinfo {author} {\bibfnamefont
  {R.}~\bibnamefont {Liang}}, \bibinfo {author} {\bibfnamefont {D.~A.}\
  \bibnamefont {Bonn}}, \bibinfo {author} {\bibfnamefont {W.~N.}\ \bibnamefont
  {Hardy}}, \bibinfo {author} {\bibfnamefont {A.}~\bibnamefont {Watenphul}},
  \bibinfo {author} {\bibfnamefont {M.}~\bibnamefont {v.~Zimmermann}}, \bibinfo
  {author} {\bibfnamefont {E.~M.}\ \bibnamefont {Forgan}},\ and\ \bibinfo
  {author} {\bibfnamefont {S.~M.}\ \bibnamefont {Hayden}},\ }\bibfield  {title}
  {\bibinfo {title} {Direct observation of competition between
  superconductivity and charge density wave order in
  {YBa}$_2${Cu}$_3${O}$_{6.67}$},\ }\href {https://doi.org/10.1038/nphys2456}
  {\bibfield  {journal} {\bibinfo  {journal} {Nat. Phys.}\ }\textbf {\bibinfo
  {volume} {8}},\ \bibinfo {pages} {871} (\bibinfo {year} {2012})}\BibitemShut
  {NoStop}%
\bibitem [{\citenamefont {da~Silva~Neto}\ \emph {et~al.}(2013)\citenamefont
  {da~Silva~Neto}, \citenamefont {Aynajian}, \citenamefont {Frano},
  \citenamefont {Comin}, \citenamefont {Schierle}, \citenamefont {Weschke},
  \citenamefont {Gyenis}, \citenamefont {Wen}, \citenamefont {Schneeloch},
  \citenamefont {Xu}, \citenamefont {Ono}, \citenamefont {Gu}, \citenamefont
  {Tacon},\ and\ \citenamefont {Yazdani}}]{SilvaNeto2013_SAFC}%
  \BibitemOpen
  \bibfield  {author} {\bibinfo {author} {\bibfnamefont {E.~H.}\ \bibnamefont
  {da~Silva~Neto}}, \bibinfo {author} {\bibfnamefont {P.}~\bibnamefont
  {Aynajian}}, \bibinfo {author} {\bibfnamefont {A.}~\bibnamefont {Frano}},
  \bibinfo {author} {\bibfnamefont {R.}~\bibnamefont {Comin}}, \bibinfo
  {author} {\bibfnamefont {E.}~\bibnamefont {Schierle}}, \bibinfo {author}
  {\bibfnamefont {E.}~\bibnamefont {Weschke}}, \bibinfo {author} {\bibfnamefont
  {A.}~\bibnamefont {Gyenis}}, \bibinfo {author} {\bibfnamefont
  {J.}~\bibnamefont {Wen}}, \bibinfo {author} {\bibfnamefont {J.}~\bibnamefont
  {Schneeloch}}, \bibinfo {author} {\bibfnamefont {Z.}~\bibnamefont {Xu}},
  \bibinfo {author} {\bibfnamefont {S.}~\bibnamefont {Ono}}, \bibinfo {author}
  {\bibfnamefont {G.}~\bibnamefont {Gu}}, \bibinfo {author} {\bibfnamefont
  {M.~L.}\ \bibnamefont {Tacon}},\ and\ \bibinfo {author} {\bibfnamefont
  {A.}~\bibnamefont {Yazdani}},\ }\bibfield  {title} {\bibinfo {title}
  {Ubiquitous interplay between charge ordering and high-temperature
  superconductivity in cuprates},\ }\href
  {https://doi.org/10.1126/science.1243479} {\bibfield  {journal} {\bibinfo
  {journal} {Science}\ }\textbf {\bibinfo {volume} {343}},\ \bibinfo {pages}
  {393} (\bibinfo {year} {2013})}\BibitemShut {NoStop}%
\bibitem [{\citenamefont {Comin}\ \emph {et~al.}(2013)\citenamefont {Comin},
  \citenamefont {Frano}, \citenamefont {Yee}, \citenamefont {Yoshida},
  \citenamefont {Eisaki}, \citenamefont {Schierle}, \citenamefont {Weschke},
  \citenamefont {Sutarto}, \citenamefont {He}, \citenamefont {Soumyanarayanan},
  \citenamefont {He}, \citenamefont {Tacon}, \citenamefont {Elfimov},
  \citenamefont {Hoffman}, \citenamefont {Sawatzky}, \citenamefont {Keimer},\
  and\ \citenamefont {Damascelli}}]{Comin2013_CFYY}%
  \BibitemOpen
  \bibfield  {author} {\bibinfo {author} {\bibfnamefont {R.}~\bibnamefont
  {Comin}}, \bibinfo {author} {\bibfnamefont {A.}~\bibnamefont {Frano}},
  \bibinfo {author} {\bibfnamefont {M.~M.}\ \bibnamefont {Yee}}, \bibinfo
  {author} {\bibfnamefont {Y.}~\bibnamefont {Yoshida}}, \bibinfo {author}
  {\bibfnamefont {H.}~\bibnamefont {Eisaki}}, \bibinfo {author} {\bibfnamefont
  {E.}~\bibnamefont {Schierle}}, \bibinfo {author} {\bibfnamefont
  {E.}~\bibnamefont {Weschke}}, \bibinfo {author} {\bibfnamefont
  {R.}~\bibnamefont {Sutarto}}, \bibinfo {author} {\bibfnamefont
  {F.}~\bibnamefont {He}}, \bibinfo {author} {\bibfnamefont {A.}~\bibnamefont
  {Soumyanarayanan}}, \bibinfo {author} {\bibfnamefont {Y.}~\bibnamefont {He}},
  \bibinfo {author} {\bibfnamefont {M.~L.}\ \bibnamefont {Tacon}}, \bibinfo
  {author} {\bibfnamefont {I.~S.}\ \bibnamefont {Elfimov}}, \bibinfo {author}
  {\bibfnamefont {J.~E.}\ \bibnamefont {Hoffman}}, \bibinfo {author}
  {\bibfnamefont {G.~A.}\ \bibnamefont {Sawatzky}}, \bibinfo {author}
  {\bibfnamefont {B.}~\bibnamefont {Keimer}},\ and\ \bibinfo {author}
  {\bibfnamefont {A.}~\bibnamefont {Damascelli}},\ }\bibfield  {title}
  {\bibinfo {title} {Charge order driven by {F}ermi-arc instability in
  {Bi}$_2${Sr}$_{2-x}${La}$_x${CuO}$_{6 + \delta}$},\ }\href
  {https://doi.org/10.1126/science.1242996} {\bibfield  {journal} {\bibinfo
  {journal} {Science}\ }\textbf {\bibinfo {volume} {343}},\ \bibinfo {pages}
  {390} (\bibinfo {year} {2013})}\BibitemShut {NoStop}%
\bibitem [{\citenamefont {Tabis}\ \emph {et~al.}(2014)\citenamefont {Tabis},
  \citenamefont {Li}, \citenamefont {Tacon}, \citenamefont {Braicovich},
  \citenamefont {Kreyssig}, \citenamefont {Minola}, \citenamefont {Dellea},
  \citenamefont {Weschke}, \citenamefont {Veit}, \citenamefont {Ramazanoglu},
  \citenamefont {Goldman}, \citenamefont {Schmitt}, \citenamefont
  {Ghiringhelli}, \citenamefont {Bari{\v{s}}i{\'{c}}}, \citenamefont {Chan},
  \citenamefont {Dorow}, \citenamefont {Yu}, \citenamefont {Zhao},
  \citenamefont {Keimer},\ and\ \citenamefont {Greven}}]{Tabis2014_TLTB}%
  \BibitemOpen
  \bibfield  {author} {\bibinfo {author} {\bibfnamefont {W.}~\bibnamefont
  {Tabis}}, \bibinfo {author} {\bibfnamefont {Y.}~\bibnamefont {Li}}, \bibinfo
  {author} {\bibfnamefont {M.~L.}\ \bibnamefont {Tacon}}, \bibinfo {author}
  {\bibfnamefont {L.}~\bibnamefont {Braicovich}}, \bibinfo {author}
  {\bibfnamefont {A.}~\bibnamefont {Kreyssig}}, \bibinfo {author}
  {\bibfnamefont {M.}~\bibnamefont {Minola}}, \bibinfo {author} {\bibfnamefont
  {G.}~\bibnamefont {Dellea}}, \bibinfo {author} {\bibfnamefont
  {E.}~\bibnamefont {Weschke}}, \bibinfo {author} {\bibfnamefont {M.~J.}\
  \bibnamefont {Veit}}, \bibinfo {author} {\bibfnamefont {M.}~\bibnamefont
  {Ramazanoglu}}, \bibinfo {author} {\bibfnamefont {A.~I.}\ \bibnamefont
  {Goldman}}, \bibinfo {author} {\bibfnamefont {T.}~\bibnamefont {Schmitt}},
  \bibinfo {author} {\bibfnamefont {G.}~\bibnamefont {Ghiringhelli}}, \bibinfo
  {author} {\bibfnamefont {N.}~\bibnamefont {Bari{\v{s}}i{\'{c}}}}, \bibinfo
  {author} {\bibfnamefont {M.~K.}\ \bibnamefont {Chan}}, \bibinfo {author}
  {\bibfnamefont {C.~J.}\ \bibnamefont {Dorow}}, \bibinfo {author}
  {\bibfnamefont {G.}~\bibnamefont {Yu}}, \bibinfo {author} {\bibfnamefont
  {X.}~\bibnamefont {Zhao}}, \bibinfo {author} {\bibfnamefont {B.}~\bibnamefont
  {Keimer}},\ and\ \bibinfo {author} {\bibfnamefont {M.}~\bibnamefont
  {Greven}},\ }\bibfield  {title} {\bibinfo {title} {Charge order and its
  connection with {F}ermi-liquid charge transport in a pristine high-${T}_c$
  cuprate},\ }\href {https://doi.org/10.1038/ncomms6875} {\bibfield  {journal}
  {\bibinfo  {journal} {Nat. Commun.}\ }\textbf {\bibinfo {volume} {5}},\
  \bibinfo {pages} {5875} (\bibinfo {year} {2014})}\BibitemShut {NoStop}%
\bibitem [{\citenamefont {Croft}\ \emph {et~al.}(2014)\citenamefont {Croft},
  \citenamefont {Lester}, \citenamefont {Senn}, \citenamefont {Bombardi},\ and\
  \citenamefont {Hayden}}]{Croft2014_CLSB}%
  \BibitemOpen
  \bibfield  {author} {\bibinfo {author} {\bibfnamefont {T.~P.}\ \bibnamefont
  {Croft}}, \bibinfo {author} {\bibfnamefont {C.}~\bibnamefont {Lester}},
  \bibinfo {author} {\bibfnamefont {M.~S.}\ \bibnamefont {Senn}}, \bibinfo
  {author} {\bibfnamefont {A.}~\bibnamefont {Bombardi}},\ and\ \bibinfo
  {author} {\bibfnamefont {S.~M.}\ \bibnamefont {Hayden}},\ }\bibfield  {title}
  {\bibinfo {title} {Charge density wave fluctuations in
  {L}a$_{2-x}${S}r$_x${C}u{O}$_4$ and their competition with
  superconductivity},\ }\href {https://doi.org/10.1103/physrevb.89.224513}
  {\bibfield  {journal} {\bibinfo  {journal} {Phys. Rev. B}\ }\textbf {\bibinfo
  {volume} {89}},\ \bibinfo {pages} {224513} (\bibinfo {year}
  {2014})}\BibitemShut {NoStop}%
\bibitem [{\citenamefont {Thampy}\ \emph {et~al.}(2014)\citenamefont {Thampy},
  \citenamefont {Dean}, \citenamefont {Christensen}, \citenamefont {Steinke},
  \citenamefont {Islam}, \citenamefont {Oda}, \citenamefont {Ido},
  \citenamefont {Momono}, \citenamefont {Wilkins},\ and\ \citenamefont
  {Hill}}]{Thampy2014_TDCS}%
  \BibitemOpen
  \bibfield  {author} {\bibinfo {author} {\bibfnamefont {V.}~\bibnamefont
  {Thampy}}, \bibinfo {author} {\bibfnamefont {M.~P.~M.}\ \bibnamefont {Dean}},
  \bibinfo {author} {\bibfnamefont {N.~B.}\ \bibnamefont {Christensen}},
  \bibinfo {author} {\bibfnamefont {L.}~\bibnamefont {Steinke}}, \bibinfo
  {author} {\bibfnamefont {Z.}~\bibnamefont {Islam}}, \bibinfo {author}
  {\bibfnamefont {M.}~\bibnamefont {Oda}}, \bibinfo {author} {\bibfnamefont
  {M.}~\bibnamefont {Ido}}, \bibinfo {author} {\bibfnamefont {N.}~\bibnamefont
  {Momono}}, \bibinfo {author} {\bibfnamefont {S.~B.}\ \bibnamefont
  {Wilkins}},\ and\ \bibinfo {author} {\bibfnamefont {J.~P.}\ \bibnamefont
  {Hill}},\ }\bibfield  {title} {\bibinfo {title} {Rotated stripe order and its
  competition with superconductivity in
  {${\mathrm{La}}_{1.88}{\mathrm{Sr}}_{0.12}{\mathrm{CuO}}_{4}$}},\ }\href
  {https://doi.org/10.1103/PhysRevB.90.100510} {\bibfield  {journal} {\bibinfo
  {journal} {Phys. Rev. B}\ }\textbf {\bibinfo {volume} {90}},\ \bibinfo
  {pages} {100510} (\bibinfo {year} {2014})}\BibitemShut {NoStop}%
\bibitem [{\citenamefont {LeBoeuf}\ \emph {et~al.}(2011)\citenamefont
  {LeBoeuf}, \citenamefont {Doiron-Leyraud}, \citenamefont {Vignolle},
  \citenamefont {Sutherland}, \citenamefont {Ramshaw}, \citenamefont
  {Levallois}, \citenamefont {Daou}, \citenamefont {Lalibert{\'{e}}},
  \citenamefont {Cyr-Choini{\`{e}}re}, \citenamefont {Chang}, \citenamefont
  {Jo}, \citenamefont {Balicas}, \citenamefont {Liang}, \citenamefont {Bonn},
  \citenamefont {Hardy}, \citenamefont {Proust},\ and\ \citenamefont
  {Taillefer}}]{LeBoeuf2011_LDVS}%
  \BibitemOpen
  \bibfield  {author} {\bibinfo {author} {\bibfnamefont {D.}~\bibnamefont
  {LeBoeuf}}, \bibinfo {author} {\bibfnamefont {N.}~\bibnamefont
  {Doiron-Leyraud}}, \bibinfo {author} {\bibfnamefont {B.}~\bibnamefont
  {Vignolle}}, \bibinfo {author} {\bibfnamefont {M.}~\bibnamefont
  {Sutherland}}, \bibinfo {author} {\bibfnamefont {B.~J.}\ \bibnamefont
  {Ramshaw}}, \bibinfo {author} {\bibfnamefont {J.}~\bibnamefont {Levallois}},
  \bibinfo {author} {\bibfnamefont {R.}~\bibnamefont {Daou}}, \bibinfo {author}
  {\bibfnamefont {F.}~\bibnamefont {Lalibert{\'{e}}}}, \bibinfo {author}
  {\bibfnamefont {O.}~\bibnamefont {Cyr-Choini{\`{e}}re}}, \bibinfo {author}
  {\bibfnamefont {J.}~\bibnamefont {Chang}}, \bibinfo {author} {\bibfnamefont
  {Y.~J.}\ \bibnamefont {Jo}}, \bibinfo {author} {\bibfnamefont
  {L.}~\bibnamefont {Balicas}}, \bibinfo {author} {\bibfnamefont
  {R.}~\bibnamefont {Liang}}, \bibinfo {author} {\bibfnamefont {D.~A.}\
  \bibnamefont {Bonn}}, \bibinfo {author} {\bibfnamefont {W.~N.}\ \bibnamefont
  {Hardy}}, \bibinfo {author} {\bibfnamefont {C.}~\bibnamefont {Proust}},\ and\
  \bibinfo {author} {\bibfnamefont {L.}~\bibnamefont {Taillefer}},\ }\bibfield
  {title} {\bibinfo {title} {Lifshitz critical point in the cuprate
  superconductor {YBa$_2$Cu$_3$O$_y$} from high-field {Hall} effect
  measurements},\ }\href {https://doi.org/10.1103/physrevb.83.054506}
  {\bibfield  {journal} {\bibinfo  {journal} {Phys. Rev. B}\ }\textbf {\bibinfo
  {volume} {83}},\ \bibinfo {pages} {054506} (\bibinfo {year}
  {2011})}\BibitemShut {NoStop}%
\bibitem [{\citenamefont {Badoux}\ \emph {et~al.}(2016)\citenamefont {Badoux},
  \citenamefont {Tabis}, \citenamefont {Lalibert{\'{e}}}, \citenamefont
  {Grissonnanche}, \citenamefont {Vignolle}, \citenamefont {Vignolles},
  \citenamefont {B{\'{e}}ard}, \citenamefont {Bonn}, \citenamefont {Hardy},
  \citenamefont {Liang}, \citenamefont {Doiron-Leyraud}, \citenamefont
  {Taillefer},\ and\ \citenamefont {Proust}}]{Badoux2016_BTLG}%
  \BibitemOpen
  \bibfield  {author} {\bibinfo {author} {\bibfnamefont {S.}~\bibnamefont
  {Badoux}}, \bibinfo {author} {\bibfnamefont {W.}~\bibnamefont {Tabis}},
  \bibinfo {author} {\bibfnamefont {F.}~\bibnamefont {Lalibert{\'{e}}}},
  \bibinfo {author} {\bibfnamefont {G.}~\bibnamefont {Grissonnanche}}, \bibinfo
  {author} {\bibfnamefont {B.}~\bibnamefont {Vignolle}}, \bibinfo {author}
  {\bibfnamefont {D.}~\bibnamefont {Vignolles}}, \bibinfo {author}
  {\bibfnamefont {J.}~\bibnamefont {B{\'{e}}ard}}, \bibinfo {author}
  {\bibfnamefont {D.~A.}\ \bibnamefont {Bonn}}, \bibinfo {author}
  {\bibfnamefont {W.~N.}\ \bibnamefont {Hardy}}, \bibinfo {author}
  {\bibfnamefont {R.}~\bibnamefont {Liang}}, \bibinfo {author} {\bibfnamefont
  {N.}~\bibnamefont {Doiron-Leyraud}}, \bibinfo {author} {\bibfnamefont
  {L.}~\bibnamefont {Taillefer}},\ and\ \bibinfo {author} {\bibfnamefont
  {C.}~\bibnamefont {Proust}},\ }\bibfield  {title} {\bibinfo {title} {Change
  of carrier density at the pseudogap critical point of a cuprate
  superconductor},\ }\href {https://doi.org/10.1038/nature16983} {\bibfield
  {journal} {\bibinfo  {journal} {Nature}\ }\textbf {\bibinfo {volume} {531}},\
  \bibinfo {pages} {210} (\bibinfo {year} {2016})}\BibitemShut {NoStop}%
\bibitem [{\citenamefont {Storey}(2016)}]{Storey2016}%
  \BibitemOpen
  \bibfield  {author} {\bibinfo {author} {\bibfnamefont {J.~G.}\ \bibnamefont
  {Storey}},\ }\bibfield  {title} {\bibinfo {title} {Hall effect and fermi
  surface reconstruction via electron pockets in the high-${T}_c$ cuprates},\
  }\href {https://doi.org/10.1209/0295-5075/113/27003} {\bibfield  {journal}
  {\bibinfo  {journal} {{EPL}}\ }\textbf {\bibinfo {volume} {113}},\ \bibinfo
  {pages} {27003} (\bibinfo {year} {2016})}\BibitemShut {NoStop}%
\bibitem [{\citenamefont {Wade}\ \emph {et~al.}(1994)\citenamefont {Wade},
  \citenamefont {Loram}, \citenamefont {Mirza}, \citenamefont {Cooper},\ and\
  \citenamefont {Tallon}}]{Wade1994_WLMC}%
  \BibitemOpen
  \bibfield  {author} {\bibinfo {author} {\bibfnamefont {J.~M.}\ \bibnamefont
  {Wade}}, \bibinfo {author} {\bibfnamefont {J.~W.}\ \bibnamefont {Loram}},
  \bibinfo {author} {\bibfnamefont {K.~A.}\ \bibnamefont {Mirza}}, \bibinfo
  {author} {\bibfnamefont {J.~R.}\ \bibnamefont {Cooper}},\ and\ \bibinfo
  {author} {\bibfnamefont {J.~L.}\ \bibnamefont {Tallon}},\ }\bibfield  {title}
  {\bibinfo {title} {Electronic specific heat of {Tl}$_2${Ba}$_2${CuO}$_{6+
  \delta}$ from {2 K} to {300 K} for {$0 \leq \delta \leq 0.1$}},\ }\href
  {https://doi.org/10.1007/bf00730408} {\bibfield  {journal} {\bibinfo
  {journal} {J. Supercond.}\ }\textbf {\bibinfo {volume} {7}},\ \bibinfo
  {pages} {261} (\bibinfo {year} {1994})}\BibitemShut {NoStop}%
\bibitem [{\citenamefont {Fujiwara}\ \emph {et~al.}(1990)\citenamefont
  {Fujiwara}, \citenamefont {Kitaoka}, \citenamefont {Asayama}, \citenamefont
  {Shimakawa}, \citenamefont {Manako},\ and\ \citenamefont
  {Kubo}}]{Fujiwara1990_FKAS}%
  \BibitemOpen
  \bibfield  {author} {\bibinfo {author} {\bibfnamefont {K.}~\bibnamefont
  {Fujiwara}}, \bibinfo {author} {\bibfnamefont {Y.}~\bibnamefont {Kitaoka}},
  \bibinfo {author} {\bibfnamefont {K.}~\bibnamefont {Asayama}}, \bibinfo
  {author} {\bibfnamefont {Y.}~\bibnamefont {Shimakawa}}, \bibinfo {author}
  {\bibfnamefont {T.}~\bibnamefont {Manako}},\ and\ \bibinfo {author}
  {\bibfnamefont {Y.}~\bibnamefont {Kubo}},\ }\bibfield  {title} {\bibinfo
  {title} {$^{63}${Cu} knight shift study in high-${T}_c$ superconductor
  {Tl}$_2${Ba}$_2${CuO}$_{6+ y}$ with a single {CuO}$_2$ layer},\ }\href
  {https://doi.org/10.1143/jpsj.59.3459} {\bibfield  {journal} {\bibinfo
  {journal} {J. Phys. Soc. Jpn.}\ }\textbf {\bibinfo {volume} {59}},\ \bibinfo
  {pages} {3459} (\bibinfo {year} {1990})}\BibitemShut {NoStop}%
\bibitem [{\citenamefont {Kambe}\ \emph {et~al.}(1993)\citenamefont {Kambe},
  \citenamefont {Yasuoka}, \citenamefont {Hayashi},\ and\ \citenamefont
  {Ueda}}]{Kambe1993_KYHU}%
  \BibitemOpen
  \bibfield  {author} {\bibinfo {author} {\bibfnamefont {S.}~\bibnamefont
  {Kambe}}, \bibinfo {author} {\bibfnamefont {H.}~\bibnamefont {Yasuoka}},
  \bibinfo {author} {\bibfnamefont {A.}~\bibnamefont {Hayashi}},\ and\ \bibinfo
  {author} {\bibfnamefont {Y.}~\bibnamefont {Ueda}},\ }\bibfield  {title}
  {\bibinfo {title} {{NMR} study of the spin dynamics in
  {Tl}$_2${Ba}$_2${CuO}$_{6+ y}$ ({${T}_c$ = 85 K})},\ }\href
  {https://doi.org/10.1103/physrevb.47.2825} {\bibfield  {journal} {\bibinfo
  {journal} {Phys. Rev. B}\ }\textbf {\bibinfo {volume} {47}},\ \bibinfo
  {pages} {2825} (\bibinfo {year} {1993})}\BibitemShut {NoStop}%
\bibitem [{\citenamefont {Putzke}\ \emph {et~al.}(2021)\citenamefont {Putzke},
  \citenamefont {Benhabib}, \citenamefont {Tabis}, \citenamefont {Ayres},
  \citenamefont {Wang}, \citenamefont {Malone}, \citenamefont {Licciardello},
  \citenamefont {Lu}, \citenamefont {Kondo}, \citenamefont {Takeuchi},
  \citenamefont {Hussey}, \citenamefont {Cooper},\ and\ \citenamefont
  {Carrington}}]{Putzke2021_PBTA}%
  \BibitemOpen
  \bibfield  {author} {\bibinfo {author} {\bibfnamefont {C.}~\bibnamefont
  {Putzke}}, \bibinfo {author} {\bibfnamefont {S.}~\bibnamefont {Benhabib}},
  \bibinfo {author} {\bibfnamefont {W.}~\bibnamefont {Tabis}}, \bibinfo
  {author} {\bibfnamefont {J.}~\bibnamefont {Ayres}}, \bibinfo {author}
  {\bibfnamefont {Z.}~\bibnamefont {Wang}}, \bibinfo {author} {\bibfnamefont
  {L.}~\bibnamefont {Malone}}, \bibinfo {author} {\bibfnamefont
  {S.}~\bibnamefont {Licciardello}}, \bibinfo {author} {\bibfnamefont
  {J.}~\bibnamefont {Lu}}, \bibinfo {author} {\bibfnamefont {T.}~\bibnamefont
  {Kondo}}, \bibinfo {author} {\bibfnamefont {T.}~\bibnamefont {Takeuchi}},
  \bibinfo {author} {\bibfnamefont {N.~E.}\ \bibnamefont {Hussey}}, \bibinfo
  {author} {\bibfnamefont {J.~R.}\ \bibnamefont {Cooper}},\ and\ \bibinfo
  {author} {\bibfnamefont {A.}~\bibnamefont {Carrington}},\ }\bibfield  {title}
  {\bibinfo {title} {Reduced {Hall} carrier density in the overdoped strange
  metal regime of cuprate superconductors},\ }\href
  {https://doi.org/10.1038/s41567-021-01197-0} {\bibfield  {journal} {\bibinfo
  {journal} {Nat. Phys.}\ }\textbf {\bibinfo {volume} {17}},\ \bibinfo {pages}
  {826} (\bibinfo {year} {2021})}\BibitemShut {NoStop}%
\bibitem [{\citenamefont {Vignolle}\ \emph {et~al.}(2008)\citenamefont
  {Vignolle}, \citenamefont {Carrington}, \citenamefont {Cooper}, \citenamefont
  {French}, \citenamefont {Mackenzie}, \citenamefont {Jaudet}, \citenamefont
  {Vignolles}, \citenamefont {Proust},\ and\ \citenamefont
  {Hussey}}]{Vignolle2008_VCCF}%
  \BibitemOpen
  \bibfield  {author} {\bibinfo {author} {\bibfnamefont {B.}~\bibnamefont
  {Vignolle}}, \bibinfo {author} {\bibfnamefont {A.}~\bibnamefont
  {Carrington}}, \bibinfo {author} {\bibfnamefont {R.~A.}\ \bibnamefont
  {Cooper}}, \bibinfo {author} {\bibfnamefont {M.~M.~J.}\ \bibnamefont
  {French}}, \bibinfo {author} {\bibfnamefont {A.~P.}\ \bibnamefont
  {Mackenzie}}, \bibinfo {author} {\bibfnamefont {C.}~\bibnamefont {Jaudet}},
  \bibinfo {author} {\bibfnamefont {D.}~\bibnamefont {Vignolles}}, \bibinfo
  {author} {\bibfnamefont {C.}~\bibnamefont {Proust}},\ and\ \bibinfo {author}
  {\bibfnamefont {N.~E.}\ \bibnamefont {Hussey}},\ }\bibfield  {title}
  {\bibinfo {title} {Quantum oscillations in an overdoped high-${T}_c$
  superconductor},\ }\href {https://doi.org/10.1038/nature07323} {\bibfield
  {journal} {\bibinfo  {journal} {Nature}\ }\textbf {\bibinfo {volume} {455}},\
  \bibinfo {pages} {952} (\bibinfo {year} {2008})}\BibitemShut {NoStop}%
\bibitem [{\citenamefont {Bangura}\ \emph {et~al.}(2010)\citenamefont
  {Bangura}, \citenamefont {Rourke}, \citenamefont {Benseman}, \citenamefont
  {Matusiak}, \citenamefont {Cooper}, \citenamefont {Hussey},\ and\
  \citenamefont {Carrington}}]{Bangura2010}%
  \BibitemOpen
  \bibfield  {author} {\bibinfo {author} {\bibfnamefont {A.~F.}\ \bibnamefont
  {Bangura}}, \bibinfo {author} {\bibfnamefont {P.~M.~C.}\ \bibnamefont
  {Rourke}}, \bibinfo {author} {\bibfnamefont {T.~M.}\ \bibnamefont
  {Benseman}}, \bibinfo {author} {\bibfnamefont {M.}~\bibnamefont {Matusiak}},
  \bibinfo {author} {\bibfnamefont {J.~R.}\ \bibnamefont {Cooper}}, \bibinfo
  {author} {\bibfnamefont {N.~E.}\ \bibnamefont {Hussey}},\ and\ \bibinfo
  {author} {\bibfnamefont {A.}~\bibnamefont {Carrington}},\ }\bibfield  {title}
  {\bibinfo {title} {Fermi surface and electronic homogeneity of the overdoped
  cuprate superconductor {Tl}$_2${Ba}$_2${CuO}$_{6+ \delta}$ as revealed by
  quantum oscillations},\ }\href {https://doi.org/10.1103/PhysRevB.82.140501}
  {\bibfield  {journal} {\bibinfo  {journal} {Phys. Rev. B}\ }\textbf {\bibinfo
  {volume} {82}},\ \bibinfo {pages} {140501} (\bibinfo {year}
  {2010})}\BibitemShut {NoStop}%
\bibitem [{\citenamefont {Rourke}\ \emph {et~al.}(2010)\citenamefont {Rourke},
  \citenamefont {Bangura}, \citenamefont {Benseman}, \citenamefont {Matusiak},
  \citenamefont {Cooper}, \citenamefont {Carrington},\ and\ \citenamefont
  {Hussey}}]{Rourke2010_RBBM}%
  \BibitemOpen
  \bibfield  {author} {\bibinfo {author} {\bibfnamefont {P.~M.~C.}\
  \bibnamefont {Rourke}}, \bibinfo {author} {\bibfnamefont {A.~F.}\
  \bibnamefont {Bangura}}, \bibinfo {author} {\bibfnamefont {T.~M.}\
  \bibnamefont {Benseman}}, \bibinfo {author} {\bibfnamefont {M.}~\bibnamefont
  {Matusiak}}, \bibinfo {author} {\bibfnamefont {J.~R.}\ \bibnamefont
  {Cooper}}, \bibinfo {author} {\bibfnamefont {A.}~\bibnamefont {Carrington}},\
  and\ \bibinfo {author} {\bibfnamefont {N.~E.}\ \bibnamefont {Hussey}},\
  }\bibfield  {title} {\bibinfo {title} {A detailed de {Haas}-van {Alphen}
  effect study of the overdoped cuprate {Tl}$_2${Ba}$_2${CuO}$_{6+ \delta}$},\
  }\href {https://doi.org/10.1088/1367-2630/12/10/105009} {\bibfield  {journal}
  {\bibinfo  {journal} {New J. Phys.}\ }\textbf {\bibinfo {volume} {12}},\
  \bibinfo {pages} {105009} (\bibinfo {year} {2010})}\BibitemShut {NoStop}%
\bibitem [{\citenamefont {Hussey}\ \emph {et~al.}(2003)\citenamefont {Hussey},
  \citenamefont {Abdel-Jawad}, \citenamefont {Carrington}, \citenamefont
  {Mackenzie},\ and\ \citenamefont {Balicas}}]{Hussey2003_HACM}%
  \BibitemOpen
  \bibfield  {author} {\bibinfo {author} {\bibfnamefont {N.~E.}\ \bibnamefont
  {Hussey}}, \bibinfo {author} {\bibfnamefont {M.}~\bibnamefont {Abdel-Jawad}},
  \bibinfo {author} {\bibfnamefont {A.}~\bibnamefont {Carrington}}, \bibinfo
  {author} {\bibfnamefont {A.~P.}\ \bibnamefont {Mackenzie}},\ and\ \bibinfo
  {author} {\bibfnamefont {L.}~\bibnamefont {Balicas}},\ }\bibfield  {title}
  {\bibinfo {title} {A coherent three-dimensional {F}ermi surface in a
  high-transition-temperature superconductor},\ }\href
  {https://doi.org/10.1038/nature01981} {\bibfield  {journal} {\bibinfo
  {journal} {Nature}\ }\textbf {\bibinfo {volume} {425}},\ \bibinfo {pages}
  {814} (\bibinfo {year} {2003})}\BibitemShut {NoStop}%
\bibitem [{\citenamefont {Abdel-Jawad}\ \emph {et~al.}(2006)\citenamefont
  {Abdel-Jawad}, \citenamefont {Kennett}, \citenamefont {Balicas},
  \citenamefont {Carrington}, \citenamefont {Mackenzie}, \citenamefont
  {McKenzie},\ and\ \citenamefont {Hussey}}]{AbdelJawad2006_AKBC}%
  \BibitemOpen
  \bibfield  {author} {\bibinfo {author} {\bibfnamefont {M.}~\bibnamefont
  {Abdel-Jawad}}, \bibinfo {author} {\bibfnamefont {M.~P.}\ \bibnamefont
  {Kennett}}, \bibinfo {author} {\bibfnamefont {L.}~\bibnamefont {Balicas}},
  \bibinfo {author} {\bibfnamefont {A.}~\bibnamefont {Carrington}}, \bibinfo
  {author} {\bibfnamefont {A.~P.}\ \bibnamefont {Mackenzie}}, \bibinfo {author}
  {\bibfnamefont {R.~H.}\ \bibnamefont {McKenzie}},\ and\ \bibinfo {author}
  {\bibfnamefont {N.~E.}\ \bibnamefont {Hussey}},\ }\bibfield  {title}
  {\bibinfo {title} {Anisotropic scattering and anomalous normal-state
  transport in a high-temperature superconductor},\ }\href
  {https://doi.org/10.1038/nphys449} {\bibfield  {journal} {\bibinfo  {journal}
  {Nat. Phys.}\ }\textbf {\bibinfo {volume} {2}},\ \bibinfo {pages} {821}
  (\bibinfo {year} {2006})}\BibitemShut {NoStop}%
\bibitem [{\citenamefont {Plat{\'{e}}}\ \emph {et~al.}(2005)\citenamefont
  {Plat{\'{e}}}, \citenamefont {Mottershead}, \citenamefont {Elfimov},
  \citenamefont {Peets}, \citenamefont {Liang}, \citenamefont {Bonn},
  \citenamefont {Hardy}, \citenamefont {Chiuzbaian}, \citenamefont {Falub},
  \citenamefont {Shi}, \citenamefont {Patthey},\ and\ \citenamefont
  {Damascelli}}]{Plate2005_PMEP}%
  \BibitemOpen
  \bibfield  {author} {\bibinfo {author} {\bibfnamefont {M.}~\bibnamefont
  {Plat{\'{e}}}}, \bibinfo {author} {\bibfnamefont {J.~D.~F.}\ \bibnamefont
  {Mottershead}}, \bibinfo {author} {\bibfnamefont {I.~S.}\ \bibnamefont
  {Elfimov}}, \bibinfo {author} {\bibfnamefont {D.~C.}\ \bibnamefont {Peets}},
  \bibinfo {author} {\bibfnamefont {R.}~\bibnamefont {Liang}}, \bibinfo
  {author} {\bibfnamefont {D.~A.}\ \bibnamefont {Bonn}}, \bibinfo {author}
  {\bibfnamefont {W.~N.}\ \bibnamefont {Hardy}}, \bibinfo {author}
  {\bibfnamefont {S.}~\bibnamefont {Chiuzbaian}}, \bibinfo {author}
  {\bibfnamefont {M.}~\bibnamefont {Falub}}, \bibinfo {author} {\bibfnamefont
  {M.}~\bibnamefont {Shi}}, \bibinfo {author} {\bibfnamefont {L.}~\bibnamefont
  {Patthey}},\ and\ \bibinfo {author} {\bibfnamefont {A.}~\bibnamefont
  {Damascelli}},\ }\bibfield  {title} {\bibinfo {title} {Fermi surface and
  quasiparticle excitations of overdoped {Tl}$_2${B}a$_2${CuO}$_{6+\delta}$},\
  }\href {https://doi.org/10.1103/physrevlett.95.077001} {\bibfield  {journal}
  {\bibinfo  {journal} {Phys. Rev. Lett.}\ }\textbf {\bibinfo {volume} {95}},\
  \bibinfo {pages} {077001} (\bibinfo {year} {2005})}\BibitemShut {NoStop}%
\bibitem [{\citenamefont {Lin}\ \emph {et~al.}(2020)\citenamefont {Lin},
  \citenamefont {Miao}, \citenamefont {Mazzone}, \citenamefont {Gu},
  \citenamefont {Nag}, \citenamefont {Walters}, \citenamefont
  {Garc{\'{\i}}a-Fern{\'{a}}ndez}, \citenamefont {Barbour}, \citenamefont
  {Pelliciari}, \citenamefont {Jarrige}, \citenamefont {Oda}, \citenamefont
  {Kurosawa}, \citenamefont {Momono}, \citenamefont {Zhou}, \citenamefont
  {Bisogni}, \citenamefont {Liu},\ and\ \citenamefont {Dean}}]{Lin2020_LMMG}%
  \BibitemOpen
  \bibfield  {author} {\bibinfo {author} {\bibfnamefont {J.}~\bibnamefont
  {Lin}}, \bibinfo {author} {\bibfnamefont {H.}~\bibnamefont {Miao}}, \bibinfo
  {author} {\bibfnamefont {D.}~\bibnamefont {Mazzone}}, \bibinfo {author}
  {\bibfnamefont {G.}~\bibnamefont {Gu}}, \bibinfo {author} {\bibfnamefont
  {A.}~\bibnamefont {Nag}}, \bibinfo {author} {\bibfnamefont {A.}~\bibnamefont
  {Walters}}, \bibinfo {author} {\bibfnamefont {M.}~\bibnamefont
  {Garc{\'{\i}}a-Fern{\'{a}}ndez}}, \bibinfo {author} {\bibfnamefont
  {A.}~\bibnamefont {Barbour}}, \bibinfo {author} {\bibfnamefont
  {J.}~\bibnamefont {Pelliciari}}, \bibinfo {author} {\bibfnamefont
  {I.}~\bibnamefont {Jarrige}}, \bibinfo {author} {\bibfnamefont
  {M.}~\bibnamefont {Oda}}, \bibinfo {author} {\bibfnamefont {K.}~\bibnamefont
  {Kurosawa}}, \bibinfo {author} {\bibfnamefont {N.}~\bibnamefont {Momono}},
  \bibinfo {author} {\bibfnamefont {K.-J.}\ \bibnamefont {Zhou}}, \bibinfo
  {author} {\bibfnamefont {V.}~\bibnamefont {Bisogni}}, \bibinfo {author}
  {\bibfnamefont {X.}~\bibnamefont {Liu}},\ and\ \bibinfo {author}
  {\bibfnamefont {M.}~\bibnamefont {Dean}},\ }\bibfield  {title} {\bibinfo
  {title} {Strongly correlated charge density wave in
  {L}a$_{2-x}${S}r$_x${C}u{O}$_4$ evidenced by doping-dependent phonon
  anomaly},\ }\href {https://doi.org/10.1103/physrevlett.124.207005} {\bibfield
   {journal} {\bibinfo  {journal} {Phys. Rev. Lett.}\ }\textbf {\bibinfo
  {volume} {124}},\ \bibinfo {pages} {207005} (\bibinfo {year}
  {2020})}\BibitemShut {NoStop}%
\bibitem [{\citenamefont {Peng}\ \emph {et~al.}(2018)\citenamefont {Peng},
  \citenamefont {Fumagalli}, \citenamefont {Ding}, \citenamefont {Minola},
  \citenamefont {Caprara}, \citenamefont {Betto}, \citenamefont {Bluschke},
  \citenamefont {Luca}, \citenamefont {Kummer}, \citenamefont
  {Lefran{\c{c}}ois}, \citenamefont {Salluzzo}, \citenamefont {Suzuki},
  \citenamefont {Tacon}, \citenamefont {Zhou}, \citenamefont {Brookes},
  \citenamefont {Keimer}, \citenamefont {Braicovich}, \citenamefont {Grilli},\
  and\ \citenamefont {Ghiringhelli}}]{Peng2018_PFDM}%
  \BibitemOpen
  \bibfield  {author} {\bibinfo {author} {\bibfnamefont {Y.~Y.}\ \bibnamefont
  {Peng}}, \bibinfo {author} {\bibfnamefont {R.}~\bibnamefont {Fumagalli}},
  \bibinfo {author} {\bibfnamefont {Y.}~\bibnamefont {Ding}}, \bibinfo {author}
  {\bibfnamefont {M.}~\bibnamefont {Minola}}, \bibinfo {author} {\bibfnamefont
  {S.}~\bibnamefont {Caprara}}, \bibinfo {author} {\bibfnamefont
  {D.}~\bibnamefont {Betto}}, \bibinfo {author} {\bibfnamefont
  {M.}~\bibnamefont {Bluschke}}, \bibinfo {author} {\bibfnamefont {G.~M.~D.}\
  \bibnamefont {Luca}}, \bibinfo {author} {\bibfnamefont {K.}~\bibnamefont
  {Kummer}}, \bibinfo {author} {\bibfnamefont {E.}~\bibnamefont
  {Lefran{\c{c}}ois}}, \bibinfo {author} {\bibfnamefont {M.}~\bibnamefont
  {Salluzzo}}, \bibinfo {author} {\bibfnamefont {H.}~\bibnamefont {Suzuki}},
  \bibinfo {author} {\bibfnamefont {M.~L.}\ \bibnamefont {Tacon}}, \bibinfo
  {author} {\bibfnamefont {X.~J.}\ \bibnamefont {Zhou}}, \bibinfo {author}
  {\bibfnamefont {N.~B.}\ \bibnamefont {Brookes}}, \bibinfo {author}
  {\bibfnamefont {B.}~\bibnamefont {Keimer}}, \bibinfo {author} {\bibfnamefont
  {L.}~\bibnamefont {Braicovich}}, \bibinfo {author} {\bibfnamefont
  {M.}~\bibnamefont {Grilli}},\ and\ \bibinfo {author} {\bibfnamefont
  {G.}~\bibnamefont {Ghiringhelli}},\ }\bibfield  {title} {\bibinfo {title}
  {Re-entrant charge order in overdoped
  ({Bi,Pb})$_{2.12}${Sr}$_{1.88}${CuO}$_{6+\delta}$ outside the pseudogap
  regime},\ }\href {https://doi.org/10.1038/s41563-018-0108-3} {\bibfield
  {journal} {\bibinfo  {journal} {Nat. Mater.}\ }\textbf {\bibinfo {volume}
  {17}},\ \bibinfo {pages} {697} (\bibinfo {year} {2018})}\BibitemShut
  {NoStop}%
\bibitem [{\citenamefont {He}\ \emph {et~al.}(2018)\citenamefont {He},
  \citenamefont {Wu}, \citenamefont {Song}, \citenamefont {Lee}, \citenamefont
  {Said}, \citenamefont {Alatas}, \citenamefont {Bosak}, \citenamefont
  {Girard}, \citenamefont {Souliou}, \citenamefont {Ruiz}, \citenamefont
  {Hepting}, \citenamefont {Bluschke}, \citenamefont {Schierle}, \citenamefont
  {Weschke}, \citenamefont {Lee}, \citenamefont {Jang}, \citenamefont {Huang},
  \citenamefont {Hashimoto}, \citenamefont {Lu}, \citenamefont {Song},
  \citenamefont {Yoshida}, \citenamefont {Eisaki}, \citenamefont {Shen},
  \citenamefont {Birgeneau}, \citenamefont {Yi},\ and\ \citenamefont
  {Frano}}]{He2018_HWSL}%
  \BibitemOpen
  \bibfield  {author} {\bibinfo {author} {\bibfnamefont {Y.}~\bibnamefont
  {He}}, \bibinfo {author} {\bibfnamefont {S.}~\bibnamefont {Wu}}, \bibinfo
  {author} {\bibfnamefont {Y.}~\bibnamefont {Song}}, \bibinfo {author}
  {\bibfnamefont {W.-S.}\ \bibnamefont {Lee}}, \bibinfo {author} {\bibfnamefont
  {A.~H.}\ \bibnamefont {Said}}, \bibinfo {author} {\bibfnamefont
  {A.}~\bibnamefont {Alatas}}, \bibinfo {author} {\bibfnamefont
  {A.}~\bibnamefont {Bosak}}, \bibinfo {author} {\bibfnamefont
  {A.}~\bibnamefont {Girard}}, \bibinfo {author} {\bibfnamefont {S.~M.}\
  \bibnamefont {Souliou}}, \bibinfo {author} {\bibfnamefont {A.}~\bibnamefont
  {Ruiz}}, \bibinfo {author} {\bibfnamefont {M.}~\bibnamefont {Hepting}},
  \bibinfo {author} {\bibfnamefont {M.}~\bibnamefont {Bluschke}}, \bibinfo
  {author} {\bibfnamefont {E.}~\bibnamefont {Schierle}}, \bibinfo {author}
  {\bibfnamefont {E.}~\bibnamefont {Weschke}}, \bibinfo {author} {\bibfnamefont
  {J.-S.}\ \bibnamefont {Lee}}, \bibinfo {author} {\bibfnamefont
  {H.}~\bibnamefont {Jang}}, \bibinfo {author} {\bibfnamefont {H.}~\bibnamefont
  {Huang}}, \bibinfo {author} {\bibfnamefont {M.}~\bibnamefont {Hashimoto}},
  \bibinfo {author} {\bibfnamefont {D.-H.}\ \bibnamefont {Lu}}, \bibinfo
  {author} {\bibfnamefont {D.}~\bibnamefont {Song}}, \bibinfo {author}
  {\bibfnamefont {Y.}~\bibnamefont {Yoshida}}, \bibinfo {author} {\bibfnamefont
  {H.}~\bibnamefont {Eisaki}}, \bibinfo {author} {\bibfnamefont {Z.-X.}\
  \bibnamefont {Shen}}, \bibinfo {author} {\bibfnamefont {R.~J.}\ \bibnamefont
  {Birgeneau}}, \bibinfo {author} {\bibfnamefont {M.}~\bibnamefont {Yi}},\ and\
  \bibinfo {author} {\bibfnamefont {A.}~\bibnamefont {Frano}},\ }\bibfield
  {title} {\bibinfo {title} {Persistent low-energy phonon broadening near the
  charge-order $q$ vector in the bilayer cuprate
  {Bi}$_2${Sr}$_2${CaCu}$_2${O}$_{8 + \delta}$},\ }\href
  {https://doi.org/10.1103/physrevb.98.035102} {\bibfield  {journal} {\bibinfo
  {journal} {Phys. Rev. B}\ }\textbf {\bibinfo {volume} {98}},\ \bibinfo
  {pages} {035102} (\bibinfo {year} {2018})}\BibitemShut {NoStop}%
\bibitem [{\citenamefont {Le~Tacon}\ \emph {et~al.}(2013)\citenamefont
  {Le~Tacon}, \citenamefont {Minola}, \citenamefont {Peets}, \citenamefont
  {Moretti~Sala}, \citenamefont {Blanco-Canosa}, \citenamefont {Hinkov},
  \citenamefont {Liang}, \citenamefont {Bonn}, \citenamefont {Hardy},
  \citenamefont {Lin}, \citenamefont {Schmitt}, \citenamefont {Braicovich},
  \citenamefont {Ghiringhelli},\ and\ \citenamefont {Keimer}}]{Tacon2013_TMPM}%
  \BibitemOpen
  \bibfield  {author} {\bibinfo {author} {\bibfnamefont {M.}~\bibnamefont
  {Le~Tacon}}, \bibinfo {author} {\bibfnamefont {M.}~\bibnamefont {Minola}},
  \bibinfo {author} {\bibfnamefont {D.~C.}\ \bibnamefont {Peets}}, \bibinfo
  {author} {\bibfnamefont {M.}~\bibnamefont {Moretti~Sala}}, \bibinfo {author}
  {\bibfnamefont {S.}~\bibnamefont {Blanco-Canosa}}, \bibinfo {author}
  {\bibfnamefont {V.}~\bibnamefont {Hinkov}}, \bibinfo {author} {\bibfnamefont
  {R.}~\bibnamefont {Liang}}, \bibinfo {author} {\bibfnamefont {D.~A.}\
  \bibnamefont {Bonn}}, \bibinfo {author} {\bibfnamefont {W.~N.}\ \bibnamefont
  {Hardy}}, \bibinfo {author} {\bibfnamefont {C.~T.}\ \bibnamefont {Lin}},
  \bibinfo {author} {\bibfnamefont {T.}~\bibnamefont {Schmitt}}, \bibinfo
  {author} {\bibfnamefont {L.}~\bibnamefont {Braicovich}}, \bibinfo {author}
  {\bibfnamefont {G.}~\bibnamefont {Ghiringhelli}},\ and\ \bibinfo {author}
  {\bibfnamefont {B.}~\bibnamefont {Keimer}},\ }\bibfield  {title} {\bibinfo
  {title} {Dispersive spin excitations in highly overdoped cuprates revealed by
  resonant inelastic x-ray scattering},\ }\href
  {https://doi.org/10.1103/PhysRevB.88.020501} {\bibfield  {journal} {\bibinfo
  {journal} {Phys. Rev. B}\ }\textbf {\bibinfo {volume} {88}},\ \bibinfo
  {pages} {020501} (\bibinfo {year} {2013})}\BibitemShut {NoStop}%
\bibitem [{\citenamefont {Devereaux}\ \emph {et~al.}(2016)\citenamefont
  {Devereaux}, \citenamefont {Shvaika}, \citenamefont {Wu}, \citenamefont
  {Wohlfeld}, \citenamefont {Jia}, \citenamefont {Wang}, \citenamefont
  {Moritz}, \citenamefont {Chaix}, \citenamefont {Lee}, \citenamefont {Shen},
  \citenamefont {Ghiringhelli},\ and\ \citenamefont
  {Braicovich}}]{Devereaux2016_DSWW}%
  \BibitemOpen
  \bibfield  {author} {\bibinfo {author} {\bibfnamefont {T.}~\bibnamefont
  {Devereaux}}, \bibinfo {author} {\bibfnamefont {A.}~\bibnamefont {Shvaika}},
  \bibinfo {author} {\bibfnamefont {K.}~\bibnamefont {Wu}}, \bibinfo {author}
  {\bibfnamefont {K.}~\bibnamefont {Wohlfeld}}, \bibinfo {author}
  {\bibfnamefont {C.}~\bibnamefont {Jia}}, \bibinfo {author} {\bibfnamefont
  {Y.}~\bibnamefont {Wang}}, \bibinfo {author} {\bibfnamefont {B.}~\bibnamefont
  {Moritz}}, \bibinfo {author} {\bibfnamefont {L.}~\bibnamefont {Chaix}},
  \bibinfo {author} {\bibfnamefont {W.-S.}\ \bibnamefont {Lee}}, \bibinfo
  {author} {\bibfnamefont {Z.-X.}\ \bibnamefont {Shen}}, \bibinfo {author}
  {\bibfnamefont {G.}~\bibnamefont {Ghiringhelli}},\ and\ \bibinfo {author}
  {\bibfnamefont {L.}~\bibnamefont {Braicovich}},\ }\bibfield  {title}
  {\bibinfo {title} {Directly characterizing the relative strength and momentum
  dependence of electron-phonon coupling using resonant inelastic x-ray
  scattering},\ }\href {https://doi.org/10.1103/physrevx.6.041019} {\bibfield
  {journal} {\bibinfo  {journal} {Phys. Rev. X}\ }\textbf {\bibinfo {volume}
  {6}},\ \bibinfo {pages} {041019} (\bibinfo {year} {2016})}\BibitemShut
  {NoStop}%
\bibitem [{\citenamefont {Chaix}\ \emph {et~al.}(2017)\citenamefont {Chaix},
  \citenamefont {Ghiringhelli}, \citenamefont {Peng}, \citenamefont
  {Hashimoto}, \citenamefont {Moritz}, \citenamefont {Kummer}, \citenamefont
  {Brookes}, \citenamefont {He}, \citenamefont {Chen}, \citenamefont {Ishida},
  \citenamefont {Yoshida}, \citenamefont {Eisaki}, \citenamefont {Salluzzo},
  \citenamefont {Braicovich}, \citenamefont {Shen}, \citenamefont {Devereaux},\
  and\ \citenamefont {Lee}}]{Chaix2017_CGPH}%
  \BibitemOpen
  \bibfield  {author} {\bibinfo {author} {\bibfnamefont {L.}~\bibnamefont
  {Chaix}}, \bibinfo {author} {\bibfnamefont {G.}~\bibnamefont {Ghiringhelli}},
  \bibinfo {author} {\bibfnamefont {Y.~Y.}\ \bibnamefont {Peng}}, \bibinfo
  {author} {\bibfnamefont {M.}~\bibnamefont {Hashimoto}}, \bibinfo {author}
  {\bibfnamefont {B.}~\bibnamefont {Moritz}}, \bibinfo {author} {\bibfnamefont
  {K.}~\bibnamefont {Kummer}}, \bibinfo {author} {\bibfnamefont {N.~B.}\
  \bibnamefont {Brookes}}, \bibinfo {author} {\bibfnamefont {Y.}~\bibnamefont
  {He}}, \bibinfo {author} {\bibfnamefont {S.}~\bibnamefont {Chen}}, \bibinfo
  {author} {\bibfnamefont {S.}~\bibnamefont {Ishida}}, \bibinfo {author}
  {\bibfnamefont {Y.}~\bibnamefont {Yoshida}}, \bibinfo {author} {\bibfnamefont
  {H.}~\bibnamefont {Eisaki}}, \bibinfo {author} {\bibfnamefont
  {M.}~\bibnamefont {Salluzzo}}, \bibinfo {author} {\bibfnamefont
  {L.}~\bibnamefont {Braicovich}}, \bibinfo {author} {\bibfnamefont {Z.-X.}\
  \bibnamefont {Shen}}, \bibinfo {author} {\bibfnamefont {T.~P.}\ \bibnamefont
  {Devereaux}},\ and\ \bibinfo {author} {\bibfnamefont {W.-S.}\ \bibnamefont
  {Lee}},\ }\bibfield  {title} {\bibinfo {title} {Dispersive charge density
  wave excitations in {B}i$_2${S}r$_2${C}a{C}u$_2${O}$_{8 + \delta}$},\ }\href
  {https://doi.org/10.1038/nphys4157} {\bibfield  {journal} {\bibinfo
  {journal} {Nat. Phys.}\ }\textbf {\bibinfo {volume} {13}},\ \bibinfo {pages}
  {952} (\bibinfo {year} {2017})}\BibitemShut {NoStop}%
\bibitem [{\citenamefont {Peng}\ \emph {et~al.}(2020)\citenamefont {Peng},
  \citenamefont {Husain}, \citenamefont {Mitrano}, \citenamefont {Sun},
  \citenamefont {Johnson}, \citenamefont {Zakrzewski}, \citenamefont
  {MacDougall}, \citenamefont {Barbour}, \citenamefont {Jarrige}, \citenamefont
  {Bisogni},\ and\ \citenamefont {Abbamonte}}]{Peng_2020}%
  \BibitemOpen
  \bibfield  {author} {\bibinfo {author} {\bibfnamefont {Y.~Y.}\ \bibnamefont
  {Peng}}, \bibinfo {author} {\bibfnamefont {A.~A.}\ \bibnamefont {Husain}},
  \bibinfo {author} {\bibfnamefont {M.}~\bibnamefont {Mitrano}}, \bibinfo
  {author} {\bibfnamefont {S.~X.-L.}\ \bibnamefont {Sun}}, \bibinfo {author}
  {\bibfnamefont {T.~A.}\ \bibnamefont {Johnson}}, \bibinfo {author}
  {\bibfnamefont {A.~V.}\ \bibnamefont {Zakrzewski}}, \bibinfo {author}
  {\bibfnamefont {G.~J.}\ \bibnamefont {MacDougall}}, \bibinfo {author}
  {\bibfnamefont {A.}~\bibnamefont {Barbour}}, \bibinfo {author} {\bibfnamefont
  {I.}~\bibnamefont {Jarrige}}, \bibinfo {author} {\bibfnamefont
  {V.}~\bibnamefont {Bisogni}},\ and\ \bibinfo {author} {\bibfnamefont
  {P.}~\bibnamefont {Abbamonte}},\ }\bibfield  {title} {\bibinfo {title}
  {Enhanced electron-phonon coupling for charge-density-wave formation in
  {La}$_{1.8-x}${Eu}$_{0.2}${Sr}$_x${CuO}$_{4+\delta}$},\ }\href
  {https://doi.org/10.1103/PhysRevLett.125.097002} {\bibfield  {journal}
  {\bibinfo  {journal} {Phys. Rev. Lett.}\ }\textbf {\bibinfo {volume} {125}},\
  \bibinfo {pages} {097002} (\bibinfo {year} {2020})}\BibitemShut {NoStop}%
\bibitem [{\citenamefont {Miao}\ \emph {et~al.}(2021)\citenamefont {Miao},
  \citenamefont {Fabbris}, \citenamefont {Koch}, \citenamefont {Mazzone},
  \citenamefont {Nelson}, \citenamefont {Acevedo-Esteves}, \citenamefont {Gu},
  \citenamefont {Li}, \citenamefont {Yilimaz}, \citenamefont {Kaznatcheev},
  \citenamefont {Vescovo}, \citenamefont {Oda}, \citenamefont {Kurosawa},
  \citenamefont {Momono}, \citenamefont {Assefa}, \citenamefont {Robinson},
  \citenamefont {Bozin}, \citenamefont {Tranquada}, \citenamefont {Johnson},\
  and\ \citenamefont {Dean}}]{Miao2021_MFKM}%
  \BibitemOpen
  \bibfield  {author} {\bibinfo {author} {\bibfnamefont {H.}~\bibnamefont
  {Miao}}, \bibinfo {author} {\bibfnamefont {G.}~\bibnamefont {Fabbris}},
  \bibinfo {author} {\bibfnamefont {R.~J.}\ \bibnamefont {Koch}}, \bibinfo
  {author} {\bibfnamefont {D.~G.}\ \bibnamefont {Mazzone}}, \bibinfo {author}
  {\bibfnamefont {C.~S.}\ \bibnamefont {Nelson}}, \bibinfo {author}
  {\bibfnamefont {R.}~\bibnamefont {Acevedo-Esteves}}, \bibinfo {author}
  {\bibfnamefont {G.~D.}\ \bibnamefont {Gu}}, \bibinfo {author} {\bibfnamefont
  {Y.}~\bibnamefont {Li}}, \bibinfo {author} {\bibfnamefont {T.}~\bibnamefont
  {Yilimaz}}, \bibinfo {author} {\bibfnamefont {K.}~\bibnamefont
  {Kaznatcheev}}, \bibinfo {author} {\bibfnamefont {E.}~\bibnamefont
  {Vescovo}}, \bibinfo {author} {\bibfnamefont {M.}~\bibnamefont {Oda}},
  \bibinfo {author} {\bibfnamefont {T.}~\bibnamefont {Kurosawa}}, \bibinfo
  {author} {\bibfnamefont {N.}~\bibnamefont {Momono}}, \bibinfo {author}
  {\bibfnamefont {T.}~\bibnamefont {Assefa}}, \bibinfo {author} {\bibfnamefont
  {I.~K.}\ \bibnamefont {Robinson}}, \bibinfo {author} {\bibfnamefont {E.~S.}\
  \bibnamefont {Bozin}}, \bibinfo {author} {\bibfnamefont {J.~M.}\ \bibnamefont
  {Tranquada}}, \bibinfo {author} {\bibfnamefont {P.~D.}\ \bibnamefont
  {Johnson}},\ and\ \bibinfo {author} {\bibfnamefont {M.~P.~M.}\ \bibnamefont
  {Dean}},\ }\bibfield  {title} {\bibinfo {title} {Charge density waves in
  cuprate superconductors beyond the critical doping},\ }\href
  {https://doi.org/10.1038/s41535-021-00327-4} {\bibfield  {journal} {\bibinfo
  {journal} {npj Quantum Mater.}\ }\textbf {\bibinfo {volume} {6}},\ \bibinfo
  {pages} {31} (\bibinfo {year} {2021})}\BibitemShut {NoStop}%
\bibitem [{\citenamefont {Arpaia}\ \emph {et~al.}(2019)\citenamefont {Arpaia},
  \citenamefont {Caprara}, \citenamefont {Fumagalli}, \citenamefont {Vecchi},
  \citenamefont {Peng}, \citenamefont {Andersson}, \citenamefont {Betto},
  \citenamefont {Luca}, \citenamefont {Brookes}, \citenamefont {Lombardi},
  \citenamefont {Salluzzo}, \citenamefont {Braicovich}, \citenamefont {Castro},
  \citenamefont {Grilli},\ and\ \citenamefont
  {Ghiringhelli}}]{Arpaia2019_ACFV}%
  \BibitemOpen
  \bibfield  {author} {\bibinfo {author} {\bibfnamefont {R.}~\bibnamefont
  {Arpaia}}, \bibinfo {author} {\bibfnamefont {S.}~\bibnamefont {Caprara}},
  \bibinfo {author} {\bibfnamefont {R.}~\bibnamefont {Fumagalli}}, \bibinfo
  {author} {\bibfnamefont {G.~D.}\ \bibnamefont {Vecchi}}, \bibinfo {author}
  {\bibfnamefont {Y.~Y.}\ \bibnamefont {Peng}}, \bibinfo {author}
  {\bibfnamefont {E.}~\bibnamefont {Andersson}}, \bibinfo {author}
  {\bibfnamefont {D.}~\bibnamefont {Betto}}, \bibinfo {author} {\bibfnamefont
  {G.~M.~D.}\ \bibnamefont {Luca}}, \bibinfo {author} {\bibfnamefont {N.~B.}\
  \bibnamefont {Brookes}}, \bibinfo {author} {\bibfnamefont {F.}~\bibnamefont
  {Lombardi}}, \bibinfo {author} {\bibfnamefont {M.}~\bibnamefont {Salluzzo}},
  \bibinfo {author} {\bibfnamefont {L.}~\bibnamefont {Braicovich}}, \bibinfo
  {author} {\bibfnamefont {C.~D.}\ \bibnamefont {Castro}}, \bibinfo {author}
  {\bibfnamefont {M.}~\bibnamefont {Grilli}},\ and\ \bibinfo {author}
  {\bibfnamefont {G.}~\bibnamefont {Ghiringhelli}},\ }\bibfield  {title}
  {\bibinfo {title} {{Dynamical charge density fluctuations pervading the phase
  diagram of a Cu-based high-${T}_c$ superconductor}},\ }\href
  {https://doi.org/10.1126/science.aav1315} {\bibfield  {journal} {\bibinfo
  {journal} {Science}\ }\textbf {\bibinfo {volume} {365}},\ \bibinfo {pages}
  {906} (\bibinfo {year} {2019})}\BibitemShut {NoStop}%
\bibitem [{\citenamefont {Mackenzie}\ \emph {et~al.}(1996)\citenamefont
  {Mackenzie}, \citenamefont {Julian}, \citenamefont {Sinclair},\ and\
  \citenamefont {Lin}}]{Mackenzie1996_MJSL}%
  \BibitemOpen
  \bibfield  {author} {\bibinfo {author} {\bibfnamefont {A.~P.}\ \bibnamefont
  {Mackenzie}}, \bibinfo {author} {\bibfnamefont {S.~R.}\ \bibnamefont
  {Julian}}, \bibinfo {author} {\bibfnamefont {D.~C.}\ \bibnamefont
  {Sinclair}},\ and\ \bibinfo {author} {\bibfnamefont {C.~T.}\ \bibnamefont
  {Lin}},\ }\bibfield  {title} {\bibinfo {title} {Normal-state magnetotransport
  in superconducting {Tl}$_2${Ba}$_2${CuO}$_{6+\delta}$ to millikelvin
  temperatures},\ }\href {https://doi.org/10.1103/physrevb.53.5848} {\bibfield
  {journal} {\bibinfo  {journal} {Phys. Rev. B}\ }\textbf {\bibinfo {volume}
  {53}},\ \bibinfo {pages} {5848} (\bibinfo {year} {1996})}\BibitemShut
  {NoStop}%
\bibitem [{\citenamefont {Johannes}\ and\ \citenamefont
  {Mazin}(2008)}]{Johannes2008_JM}%
  \BibitemOpen
  \bibfield  {author} {\bibinfo {author} {\bibfnamefont {M.~D.}\ \bibnamefont
  {Johannes}}\ and\ \bibinfo {author} {\bibfnamefont {I.~I.}\ \bibnamefont
  {Mazin}},\ }\bibfield  {title} {\bibinfo {title} {Fermi surface nesting and
  the origin of charge density waves in metals},\ }\href
  {https://doi.org/10.1103/PhysRevB.77.165135} {\bibfield  {journal} {\bibinfo
  {journal} {Phys. Rev. B}\ }\textbf {\bibinfo {volume} {77}},\ \bibinfo
  {pages} {165135} (\bibinfo {year} {2008})}\BibitemShut {NoStop}%
\bibitem [{\citenamefont {P{\'{e}}pin}\ \emph {et~al.}(2014)\citenamefont
  {P{\'{e}}pin}, \citenamefont {de~Carvalho}, \citenamefont {Kloss},\ and\
  \citenamefont {Montiel}}]{Pepin2014_PCKM}%
  \BibitemOpen
  \bibfield  {author} {\bibinfo {author} {\bibfnamefont {C.}~\bibnamefont
  {P{\'{e}}pin}}, \bibinfo {author} {\bibfnamefont {V.~S.}\ \bibnamefont
  {de~Carvalho}}, \bibinfo {author} {\bibfnamefont {T.}~\bibnamefont {Kloss}},\
  and\ \bibinfo {author} {\bibfnamefont {X.}~\bibnamefont {Montiel}},\
  }\bibfield  {title} {\bibinfo {title} {Pseudogap, charge order, and pairing
  density wave at the hot spots in cuprate superconductors},\ }\href
  {https://doi.org/10.1103/physrevb.90.195207} {\bibfield  {journal} {\bibinfo
  {journal} {Phys. Rev. B}\ }\textbf {\bibinfo {volume} {90}},\ \bibinfo
  {pages} {195207} (\bibinfo {year} {2014})}\BibitemShut {NoStop}%
\bibitem [{\citenamefont {Kivelson}\ \emph {et~al.}(2003)\citenamefont
  {Kivelson}, \citenamefont {Bindloss}, \citenamefont {Fradkin}, \citenamefont
  {Oganesyan}, \citenamefont {Tranquada}, \citenamefont {Kapitulnik},\ and\
  \citenamefont {Howald}}]{Kivelson2003_KBFO}%
  \BibitemOpen
  \bibfield  {author} {\bibinfo {author} {\bibfnamefont {S.~A.}\ \bibnamefont
  {Kivelson}}, \bibinfo {author} {\bibfnamefont {I.~P.}\ \bibnamefont
  {Bindloss}}, \bibinfo {author} {\bibfnamefont {E.}~\bibnamefont {Fradkin}},
  \bibinfo {author} {\bibfnamefont {V.}~\bibnamefont {Oganesyan}}, \bibinfo
  {author} {\bibfnamefont {J.~M.}\ \bibnamefont {Tranquada}}, \bibinfo {author}
  {\bibfnamefont {A.}~\bibnamefont {Kapitulnik}},\ and\ \bibinfo {author}
  {\bibfnamefont {C.}~\bibnamefont {Howald}},\ }\bibfield  {title} {\bibinfo
  {title} {How to detect fluctuating stripes in the high-temperature
  superconductors},\ }\href {https://doi.org/10.1103/RevModPhys.75.1201}
  {\bibfield  {journal} {\bibinfo  {journal} {Rev. Mod. Phys.}\ }\textbf
  {\bibinfo {volume} {75}},\ \bibinfo {pages} {1201} (\bibinfo {year}
  {2003})}\BibitemShut {NoStop}%
\bibitem [{Car()}]{Carrington2021}%
  \BibitemOpen
  \href@noop {} {}\bibinfo {note} {A. Carrington (unpublished).}\BibitemShut
  {Stop}%
\bibitem [{\citenamefont {Allais}\ \emph {et~al.}(2014)\citenamefont {Allais},
  \citenamefont {Chowdhury},\ and\ \citenamefont {Sachdev}}]{Allais2014_AlCS}%
  \BibitemOpen
  \bibfield  {author} {\bibinfo {author} {\bibfnamefont {A.}~\bibnamefont
  {Allais}}, \bibinfo {author} {\bibfnamefont {D.}~\bibnamefont {Chowdhury}},\
  and\ \bibinfo {author} {\bibfnamefont {S.}~\bibnamefont {Sachdev}},\
  }\bibfield  {title} {\bibinfo {title} {Connecting high-field quantum
  oscillations to zero-field electron spectral functions in the underdoped
  cuprates},\ }\href {https://doi.org/10.1038/ncomms6771} {\bibfield  {journal}
  {\bibinfo  {journal} {Nat. Comms.}\ }\textbf {\bibinfo {volume} {5}},\
  \bibinfo {pages} {5771} (\bibinfo {year} {2014})}\BibitemShut {NoStop}%
\bibitem [{\citenamefont {Gannot}\ \emph {et~al.}(2019)\citenamefont {Gannot},
  \citenamefont {Ramshaw},\ and\ \citenamefont {Kivelson}}]{Gannot2019_GRK}%
  \BibitemOpen
  \bibfield  {author} {\bibinfo {author} {\bibfnamefont {Y.}~\bibnamefont
  {Gannot}}, \bibinfo {author} {\bibfnamefont {B.~J.}\ \bibnamefont
  {Ramshaw}},\ and\ \bibinfo {author} {\bibfnamefont {S.~A.}\ \bibnamefont
  {Kivelson}},\ }\bibfield  {title} {\bibinfo {title} {Fermi surface
  reconstruction by a charge density wave with finite correlation length},\
  }\href {https://doi.org/10.1103/PhysRevB.100.045128} {\bibfield  {journal}
  {\bibinfo  {journal} {Phys. Rev. B}\ }\textbf {\bibinfo {volume} {100}},\
  \bibinfo {pages} {045128} (\bibinfo {year} {2019})}\BibitemShut {NoStop}%
\bibitem [{\citenamefont {Lipscombe}\ \emph {et~al.}(2007)\citenamefont
  {Lipscombe}, \citenamefont {Hayden}, \citenamefont {Vignolle}, \citenamefont
  {McMorrow},\ and\ \citenamefont {Perring}}]{Lipscombe2007_LHVM}%
  \BibitemOpen
  \bibfield  {author} {\bibinfo {author} {\bibfnamefont {O.~J.}\ \bibnamefont
  {Lipscombe}}, \bibinfo {author} {\bibfnamefont {S.~M.}\ \bibnamefont
  {Hayden}}, \bibinfo {author} {\bibfnamefont {B.}~\bibnamefont {Vignolle}},
  \bibinfo {author} {\bibfnamefont {D.~F.}\ \bibnamefont {McMorrow}},\ and\
  \bibinfo {author} {\bibfnamefont {T.~G.}\ \bibnamefont {Perring}},\
  }\bibfield  {title} {\bibinfo {title} {Persistence of high-frequency spin
  fluctuations in overdoped superconducting {La$_{2-x}$Sr$_{x}$CuO$_4$
  ($x=0.22$)}},\ }\href {https://doi.org/10.1103/physrevlett.99.067002}
  {\bibfield  {journal} {\bibinfo  {journal} {Phys. Rev. Lett.}\ }\textbf
  {\bibinfo {volume} {99}},\ \bibinfo {pages} {067002} (\bibinfo {year}
  {2007})}\BibitemShut {NoStop}%
\bibitem [{\citenamefont {Hussey}(2008)}]{Hussey2008}%
  \BibitemOpen
  \bibfield  {author} {\bibinfo {author} {\bibfnamefont {N.~E.}\ \bibnamefont
  {Hussey}},\ }\bibfield  {title} {\bibinfo {title} {Phenomenology of the
  normal state in-plane transport properties of high-${T}_c$ cuprates},\ }\href
  {https://doi.org/10.1088/0953-8984/20/12/123201} {\bibfield  {journal}
  {\bibinfo  {journal} {J. Phys.: Condens. Matter}\ }\textbf {\bibinfo {volume}
  {20}},\ \bibinfo {pages} {123201} (\bibinfo {year} {2008})}\BibitemShut
  {NoStop}%
\bibitem [{\citenamefont {Ayres}\ \emph {et~al.}(2021)\citenamefont {Ayres},
  \citenamefont {Berben}, \citenamefont {{\v{C}}ulo}, \citenamefont {Hsu},
  \citenamefont {van Heumen}, \citenamefont {Huang}, \citenamefont {Zaanen},
  \citenamefont {Kondo}, \citenamefont {Takeuchi}, \citenamefont {Cooper},
  \citenamefont {Putzke}, \citenamefont {Friedemann}, \citenamefont
  {Carrington},\ and\ \citenamefont {Hussey}}]{Ayres2021}%
  \BibitemOpen
  \bibfield  {author} {\bibinfo {author} {\bibfnamefont {J.}~\bibnamefont
  {Ayres}}, \bibinfo {author} {\bibfnamefont {M.}~\bibnamefont {Berben}},
  \bibinfo {author} {\bibfnamefont {M.}~\bibnamefont {{\v{C}}ulo}}, \bibinfo
  {author} {\bibfnamefont {Y.-T.}\ \bibnamefont {Hsu}}, \bibinfo {author}
  {\bibfnamefont {E.}~\bibnamefont {van Heumen}}, \bibinfo {author}
  {\bibfnamefont {Y.}~\bibnamefont {Huang}}, \bibinfo {author} {\bibfnamefont
  {J.}~\bibnamefont {Zaanen}}, \bibinfo {author} {\bibfnamefont
  {T.}~\bibnamefont {Kondo}}, \bibinfo {author} {\bibfnamefont
  {T.}~\bibnamefont {Takeuchi}}, \bibinfo {author} {\bibfnamefont {J.~R.}\
  \bibnamefont {Cooper}}, \bibinfo {author} {\bibfnamefont {C.}~\bibnamefont
  {Putzke}}, \bibinfo {author} {\bibfnamefont {S.}~\bibnamefont {Friedemann}},
  \bibinfo {author} {\bibfnamefont {A.}~\bibnamefont {Carrington}},\ and\
  \bibinfo {author} {\bibfnamefont {N.~E.}\ \bibnamefont {Hussey}},\ }\bibfield
   {title} {\bibinfo {title} {Incoherent transport across the strange-metal
  regime of overdoped cuprates},\ }\href
  {https://doi.org/10.1038/s41586-021-03622-z} {\bibfield  {journal} {\bibinfo
  {journal} {Nature}\ }\textbf {\bibinfo {volume} {595}},\ \bibinfo {pages}
  {661} (\bibinfo {year} {2021})}\BibitemShut {NoStop}%
\bibitem [{\citenamefont {Tyler}(1998)}]{Tyler1998_Tyle}%
  \BibitemOpen
  \bibfield  {author} {\bibinfo {author} {\bibfnamefont {A.~W.}\ \bibnamefont
  {Tyler}},\ }\emph {\bibinfo {title} {An Investigation into the
  Magnetotransport Properties of Layered superconducting perovskites}},\
  \href@noop {} {Ph.D. thesis},\ \bibinfo  {school} {University of Cambridge}
  (\bibinfo {year} {1998})\BibitemShut {NoStop}%
\bibitem [{\citenamefont {Harrison}(2011)}]{Harrison2011}%
  \BibitemOpen
  \bibfield  {author} {\bibinfo {author} {\bibfnamefont {N.}~\bibnamefont
  {Harrison}},\ }\bibfield  {title} {\bibinfo {title} {Near doping-independent
  pocket area from an antinodal fermi surface instability in underdoped high
  temperature superconductors},\ }\href
  {https://doi.org/10.1103/PhysRevLett.107.186408} {\bibfield  {journal}
  {\bibinfo  {journal} {Phys. Rev. Lett.}\ }\textbf {\bibinfo {volume} {107}},\
  \bibinfo {pages} {186408} (\bibinfo {year} {2011})}\BibitemShut {NoStop}%
\bibitem [{\citenamefont {Ong}(1991)}]{Ong1991_Ong}%
  \BibitemOpen
  \bibfield  {author} {\bibinfo {author} {\bibfnamefont {N.~P.}\ \bibnamefont
  {Ong}},\ }\bibfield  {title} {\bibinfo {title} {Geometric interpretation of
  the weak-field {Hall} conductivity in two-dimensional metals with arbitrary
  fermi surface},\ }\href {https://doi.org/10.1103/physrevb.43.193} {\bibfield
  {journal} {\bibinfo  {journal} {Phys. Rev. B}\ }\textbf {\bibinfo {volume}
  {43}},\ \bibinfo {pages} {193} (\bibinfo {year} {1991})}\BibitemShut
  {NoStop}%
\end{thebibliography}
\end{document}